\pdfoutput=1
\documentclass[aps,longbibliography,superscriptaddress,eqsecnum,twocolumn,nofootinbib,subfig]{revtex4-2}
\usepackage{graphicx}
\usepackage{stmaryrd}
\usepackage{amsmath, amsfonts,amssymb}
\usepackage{mathrsfs}
\usepackage{braket}
\usepackage{xcolor}
\usepackage{xspace}
\usepackage{amscd}
\usepackage{comment}
\usepackage{bm}
\usepackage{bbm}
\definecolor{lightblue}{rgb}{0.13, 0.26, 0.99}

\usepackage[
colorlinks=true,
urlcolor=blue,
citecolor=blue,
linkcolor=blue,
hyperfootnotes=false]{hyperref}

\usepackage{appendix}

\allowdisplaybreaks

\usepackage{color}
\usepackage[normalem]{ulem}

\begin{document}

\title{
Evidence for interior-gap pair-density-wave state in Kondo-Heisenberg chains
}

\author{Yuto Hirose}
\affiliation{Quantum Matter Program, Graduate School of Advanced Science and Engineering, Hiroshima University, Higashihiroshima, Hiroshima 739-8530, Japan
}

\author{Shunsuke C. Furuya}
\affiliation{Department of Liberal Arts, Saitama Medical University, Moroyama, Saitama 350-0495, Japan}
\affiliation{Institute for Solid State Physics, The University of Tokyo, Kashiwa, 277-8581, Japan}

\author{Yasuhiro Tada}
\email[]{ytada@hiroshima-u.ac.jp}
\affiliation{Quantum Matter Program, Graduate School of Advanced Science and Engineering, Hiroshima University, Higashihiroshima, Hiroshima 739-8530, Japan
}
\affiliation{Institute for Solid State Physics, The University of Tokyo, Kashiwa, 277-8581, Japan}



\begin{abstract}
Interior-gap superconductivity has long been discussed as an exotic paired state in the presence of Fermi-surface mismatch, but its realization in canonical strongly correlated models has remained elusive. Here we present evidence that the superconducting phase of one-dimensional Kondo-Heisenberg models realizes an interior-gap pair-density-wave (PDW) state generated by strong correlations. Combining infinite density-matrix-renormalization-group (iDMRG) and finite DMRG calculations for $S=1/2$ and $S=3/2$ chains, we show that the PDW correlation is the dominant bulk superconducting correlation in the spin-gapped regime and that the momentum distribution function $n(k)$ exhibits a reconstructed structure characteristic of interior-gap physics. In particular, while the feature in $n(k)$ for the $S=1/2$ chain is only hump-like, the corresponding structure in the $S=3/2$ chain develops into a clear dip, strongly supporting the interpretation in terms of an interior-gap-like dip structure. Unlike conventional interior-gap scenarios based on a mismatch between preexisting Fermi surfaces, the present system starts from a single bare conduction-electron Fermi surface, and the additional low-energy single-particle structure emerges dynamically together with the dominant PDW correlation through the Kondo coupling. Finite DMRG data further demonstrate that boundary effects can substantially modify real-space correlations in this gapless one-dimensional system, making a direct thermodynamic-limit calculation essential for identifying the intrinsic bulk momentum structure and the dominant correlation channel.
\end{abstract}

\maketitle

\section{Introduction}
\label{sec:introduction}
Interior-gap superconductivity, or more broadly the breached-pair state, has long attracted interest as an exotic paired state realized under Fermi-surface mismatch~\cite{Sarma1963,Liu2003,Wu2003,Bedaque2003,Gubankova2003,Forbes2005,Gubbels2013,Liu2022}. 
In such a state, pairing coexists with gapless Fermi-surface-like structure, leading to a single-particle momentum distribution qualitatively different from that of an ordinary uniform BCS superconductor. At the same time, simple pairing-interaction models often find such states to be unstable, for example against phase separation or finite-momentum pairing states. It therefore remains an open question whether interior-gap-like superconductivity can be realized in canonical strongly correlated electron models.

From this perspective, the one-dimensional Kondo-Heisenberg model provides a particularly interesting setting~\cite{Affleck1996,Affleck1997,Zachar2001,Tsvelik2001,Berg2010,Fradkin2014,Fradkin2020,Moukouri1996,Edelstein2011,Liu2024,Yang2025}. It was shown that a spin-gapped phase exists over a broad parameter range in a one-dimensional Kondo lattice supplemented by a Heisenberg interaction among localized spins~\cite{Affleck1996,Affleck1997}. It was subsequently argued that this spin-gapped phase exhibits strong superconducting correlations at a finite wavevector and interpreted it as a pair-density-wave (PDW) state~\cite{Berg2010,Fradkin2014,Fradkin2020,Fradkin2015,Agterberg2020}. More recently, a reconstructed single-particle structure was reported in the same regime, including a dispersion with two minima and four Fermi points, as well as in-gap states with weight concentrated in hole pockets~\cite{Yang2025}. These results strongly suggest that the superconducting state of the Kondo-Heisenberg chain possesses a nontrivial momentum structure beyond the conventional Luther-Emery-liquid picture. More generally, pocket-like quasiparticle structures have also been discussed in the PDW literature at the mean-field level~\cite{Baruch2008,Berg2009prb}. In that broader sense, the present interior-gap interpretation may be viewed as the strongly correlated one-dimensional counterpart of a gapless pocket structure associated with finite-momentum pairing.

A key point of the present problem is that this reconstructed structure is qualitatively different from the conventional interior-gap scenario. In the latter, one usually starts from two mismatched preexisting Fermi surfaces, and the central issue is whether pairing can occur without fully removing the gapless pockets~\cite{Sarma1963,Liu2003,Wu2003,Bedaque2003,Gubankova2003,Forbes2005,Gubbels2013,Liu2022}. In the Kondo-Heisenberg chain studied here, by contrast, the bare conduction electrons have only a single Fermi surface. The additional low-energy single-particle structure is not externally given but emerges dynamically through the Kondo coupling in the presence of the Heisenberg exchange. Correspondingly, it is more natural to regard the dominant PDW correlation and the interior-gap-like single-particle reconstruction not as a simple cause-and-effect pair, but as two intertwined manifestations of the same interaction-driven correlated phase.

At the same time, the numerical identification of the bulk physics in this phase requires care because of boundary effects intrinsic to gapless one-dimensional systems. In general, open boundaries necessarily induce Friedel oscillations in the particle density, and depending on the observable, their influence can persist to long distances. In the Kondo-Heisenberg chain, finite-system density-matrix-renormalization-group (DMRG) data can therefore show substantial boundary-induced modulations in several competing correlation channels. As a result, it is generally subtle to determine from finite open chains alone which correlation---PDW, composite order, or charge correlations---is truly the most dominant in the asymptotic bulk limit. For this reason, in the present work we compute the correlation functions using infinite DMRG directly in the thermodynamic limit, and use finite DMRG as complementary information for boundary-induced real-space structure.

In this paper, by comparing the $S=1/2$ and $S=3/2$ Kondo-Heisenberg chains, we show that the spin-gapped superconducting phase exhibits an interior-gap momentum structure together with dominant PDW correlations. In particular, for $S=1/2$ the feature in the momentum distribution function $n(k)$ appears only hump-like and its interpretation remains somewhat ambiguous when viewed in isolation, whereas for $S=3/2$ the same structure develops into a clear dip. This systematic evolution strongly supports the view that the four-Fermi-point structure reported previously is naturally understood as an interior-gap reconstruction~\cite{Yang2025}. We also show, using iDMRG, that at least in the parameter regime studied here the PDW correlation is the most dominant one in the bulk. Our results therefore sharpen the existing PDW picture of the spin-gapped Kondo-Heisenberg chain in the thermodynamic limit and place its single-particle momentum structure in the broader context of interaction-generated interior-gap superconductivity.

Our results also make explicit the distinction between the Yamanaka-Oshikawa-Affleck (YOA) momentum constraint and the single-particle Fermi surface in the spin-$S$ Kondo-Heisenberg model~\cite{Yamanaka1997}: in the present state, the YOA-required large wavevector appears in the charge sector, but not as a conventional single-particle large Fermi surface.

The remainder of this paper is organized as follows. We first introduce the model and numerical methods. We then present the bulk correlation functions obtained by iDMRG and compare the $S=1/2$ and $S=3/2$ cases, identifying the dominant order in the thermodynamic limit and focusing on the emergence of the interior-gap dip structure. Next, we discuss real-space correlations and boundary-induced structures obtained from finite DMRG, thereby illustrating why a thermodynamic-limit treatment is essential in the present problem. Finally, we summarize our numerical results.

\section{model and calculation method}
\label{sec:model}
\begin{figure}[tb]
    \centering
    \includegraphics[width=0.8\linewidth]{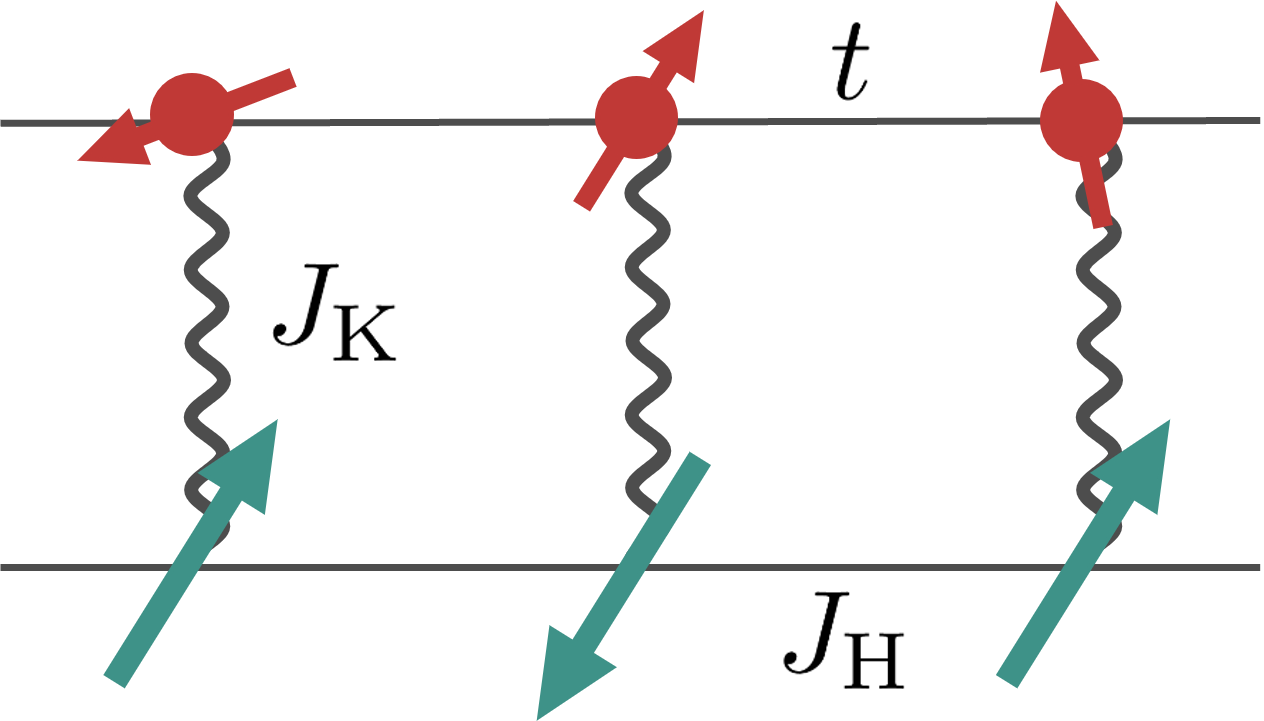}
    \caption{Schematic picture of the Kondo-Heisenberg model (\ref{eq:KHC}) with the hopping amplitude \(t\), 
  the Kondo coupling \(J_{\text{K}}\), and the exchange interaction \(J_{\text{H}}\) between localized spins.}
  \label{fig:KHC}
\end{figure}

We introduce the spin-\(S\) Kondo-Heisenberg model,
\begin{eqnarray}\label{eq:KHC}
\cal H _{\text{KH}} =&-t \sum_{i, \alpha} ( c^{\dagger}_{i \alpha} c_{i+1 \alpha} + \text{h.c.} ) + J_{\text{K}} \sum_i \bm S_i \cdot \bm s_i \nonumber\\
&+ J_{\text{H}} \sum_i \bm S_i \cdot \bm S_{i+1},
\end{eqnarray}
where \(c^{\dagger}_{i \alpha}\) (\(c_{i \alpha}\)) represents the creation (annihilation) operator of conduction electrons with spin \(\alpha = \uparrow, \downarrow\) at site-\(i\).
\(\bm S_i\) and \(\bm s_i\) represent the spin operator of localized spins and that of conduction electrons, \(\sum_{\alpha \beta} \frac{1}{2}c^{\dagger}_{i \alpha} \bm \sigma_{\alpha \beta} c_{i \beta}\), respectively.
Here, \(\bm \sigma_{\alpha \beta}\) is the Pauli matrix. 
The first term 
of Eq.~\eqref{eq:KHC}
is a hopping term of the conduction electron with a hopping amplitude \(t\). Here, \(\text{h.c.}\) means a Hermitian conjugate of the preceding term. The second term describes an antiferromagnetic exchange interaction (Kondo coupling) between \(\bm S_i\) and \(\bm s_i\). The third term describes a nearest-neighbor antiferromagnetic exchange interaction between the localized spins. The schematic picture of the model is shown in Fig.~\hyperref[fig:KHC]{\ref*{fig:KHC}}.

In this work we consider not only the $S=1/2$ Kondo-Heisenberg chain but also the $S=3/2$ case. Although one might naively regard $S=1$ as the next spin value to study, our choice is instead guided by the YOA argument~\cite{Yamanaka1997}. Their variational state carries momentum with respect to the ground-state one,
\begin{align}
P_{\mathrm{YOA}} = 2k_F + 2\pi S \quad (\mathrm{mod}\ 2\pi),
\end{align}
where $k_F=\pi n/2$ is the Fermi wavevector of the bare conduction electrons at $J_K=0$ with the filling $n$.
For half-odd-integer $S$, this gives the nontrivial momentum $2k_F+\pi$, while for integer $S$ it reduces to $2k_F$ modulo $2\pi$. Therefore, $S=3/2$ is the natural extension of the $S=1/2$ case if one wishes to preserve the same nontrivial YOA wavevector structure, whereas $S=1$ belongs to a different class in this respect~\cite{Totsuka2022}. 
Note that, at the filling $n=7/8$ which we focus on in this study, the Fermi wavevector of the conduction electrons at $J_K=0$ is $k_F=7\pi/16\simeq 0.44\pi$ and $P_{\rm YOA}=2k_F+\pi= 15\pi/8=1.875\pi=-0.125\pi$ (mod $2\pi$). 
Although this momentum is often associated with a large-Fermi-surface interpretation, we will show below that its realization in the present spin-gapped phase is more subtle.

Previous numerical studies of the Kondo-Heisenberg chain have mainly relied on finite-size DMRG calculations~\cite{Affleck1996,Affleck1997,Berg2010,Fradkin2014,Fradkin2020,Moukouri1996,Edelstein2011,Yang2025}. While this is often sufficient for identifying qualitative features, boundary effects can in general be non-negligible in gapless one-dimensional systems, as discussed in the Introduction. In particular, open boundaries necessarily induce Friedel oscillations in the local density, and depending on the observable, such boundary-induced modulations can strongly affect the apparent long-distance behavior of correlation functions. As we show below, the Kondo-Heisenberg model indeed exhibits substantial boundary effects for some observables.

To overcome this difficulty and to characterize the PDW state as an intrinsic bulk phase, we compute the momentum distribution function and bulk correlation properties using infinite DMRG (iDMRG), which directly accesses the thermodynamic limit without physical edges. This is particularly important for the present problem, where the identification of the dominant correlation channel and the subtle hump/dip structure in 
$n(k)$ require a clear separation of bulk behavior from boundary-induced distortions.
In addition, we also perform finite DMRG calculations on open chains. These data serve a complementary purpose: rather than being used to identify the bulk phase by themselves, they allow us to explicitly reveal how boundary effects appear in real-space observables and correlation functions. The combination of iDMRG and finite DMRG thus enables us to distinguish intrinsic bulk properties from boundary-induced structures in a controlled manner.

We calculate the ground state properties in the Kondo-Heisenberg model \eqref{eq:KHC} using both infinite and finite DMRG methods with TeNPy Library \cite{White1992,TeNPy2018}. We focus on the sector at the filling $n=7/8$ and \( \braket{T^z_{\rm tot}}= \sum_j \braket{S^z_j} + \braket{s^z_j} = 0 \). The coupling strength is fixed as a typical set of values for the PDW state, $J_K=J_H=2t$~\cite{Berg2010,Fradkin2020,Yang2025}, throughout this study. The open boundary condition (OBC) is imposed for finite DMRG, while there is no physical boundary for iDMRG. The bond dimension used in this study is \(\chi_{}=2000 \sim 5000\) for finite DMRG and \(\chi_{} \le 11000\) for iDMRG. 

\section{numerical results}
\label{sec:results}
In this section, we present our numerical results for the Kondo-Heisenberg chains. We begin with iDMRG calculations, which directly probe the bulk properties in the thermodynamic limit and allow us to identify the intrinsic momentum structure and dominant correlation channel of the spin-gapped phase. We then examine finite DMRG results on open chains, focusing on the boundary-induced structures that appear in real-space observables and correlations.

\subsection{Bulk properties in the thermodynamic limit}
\label{sec:bulk}
We first discuss the correlation functions for various operators in the infinite systems and identify which order is most dominant as a quasi-long-range order. For the spin sector, we can confirm that correlation functions decay exponentially, which implies existence of a spin gap. This is consistent with the previous studies~\cite{Affleck1996,Affleck1997}.
For the charge sector, we consider the charge-density-wave (CDW), charge-$2e$ pairing, charge-$4e$ pairing, composite pairing, and bond-pairing orders.
The corresponding operators are respectively defined as
\begin{align}
\mathcal{O}_{{\rm CDW}}(j)&=n_j -\braket{n_j},\\
\mathcal{O}^\dagger_{2e}(j)&=c^\dagger_{j\uparrow}c^\dagger_{j\downarrow},\\
\mathcal{O}^\dagger_{4e}(j)&=c^\dagger_{j\uparrow}c^\dagger_{j\downarrow}c^\dagger_{j+1\uparrow}c^\dagger_{j+1\downarrow},\\
\mathcal{O}^\dagger_{c}(j)&=\left[c^\dagger_{j-1\alpha}(i\bm\sigma \sigma^y)_{\alpha\beta}c^\dagger_{j+1\beta}\right]\cdot \bm S_j,\\
\mathcal{O}^\dagger_{\rm B}(j)&= \frac12 \left(c^\dagger_{j\uparrow}c^\dagger_{j+1\downarrow} - c^\dagger_{j\downarrow}c^\dagger_{j+1\uparrow} \right).
\end{align}
The bond-pairing $\mathcal{O}_{\rm B}(j)$ is even under the spatial inversion $j\leftrightarrow j+1$ and is spin-singlet pairing.  The composite operator $\mathcal{O}_c(j)$ consists of spin-triplet pairing which is odd under the inversion $j-1\leftrightarrow j+1$.
These pairing operators have been discussed in the previous studies~\cite{Zachar2001,Tsvelik2001,Berg2010}, while the CDW operator has not been numerically analyzed before to our best knowledge.
However, it is important to consider the CDW order and the pairing orders on an equal footing, since CDW state is a natural candidate for the charge sector and also a PDW state is typically considered to be an intertwined state where it is accompanied by a charge- or spin-density order~\cite{Agterberg2020}. Note that the CDW order corresponding to the discrete translation symmetry breaking can be a true long-range order even in one dimension, but it is also possible that it is the dominant quasi-long-range order over other candidates. 

\begin{figure}[htb]
\centering
    \includegraphics[width=0.8\columnwidth]{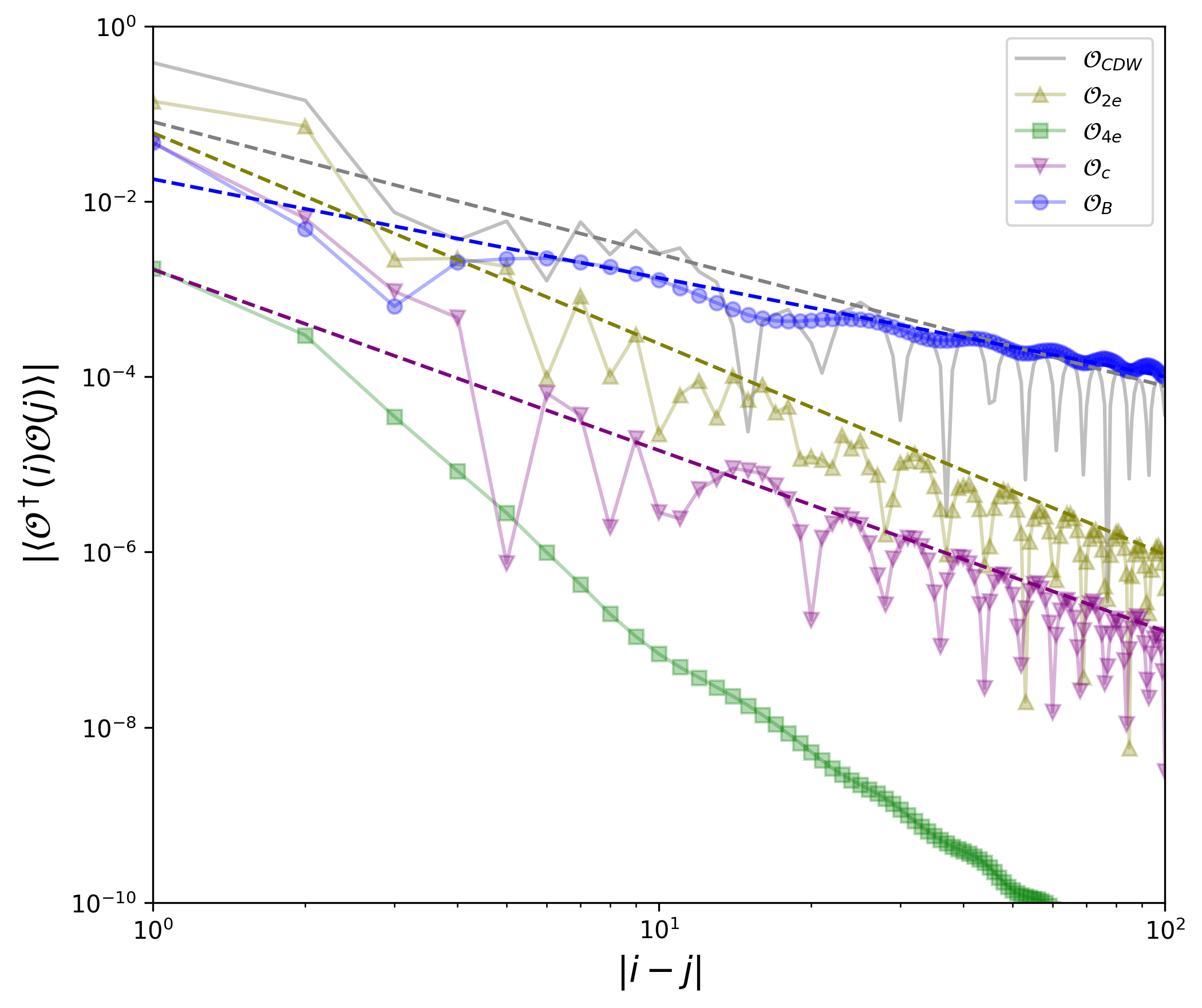}
    \includegraphics[width=0.8\columnwidth]{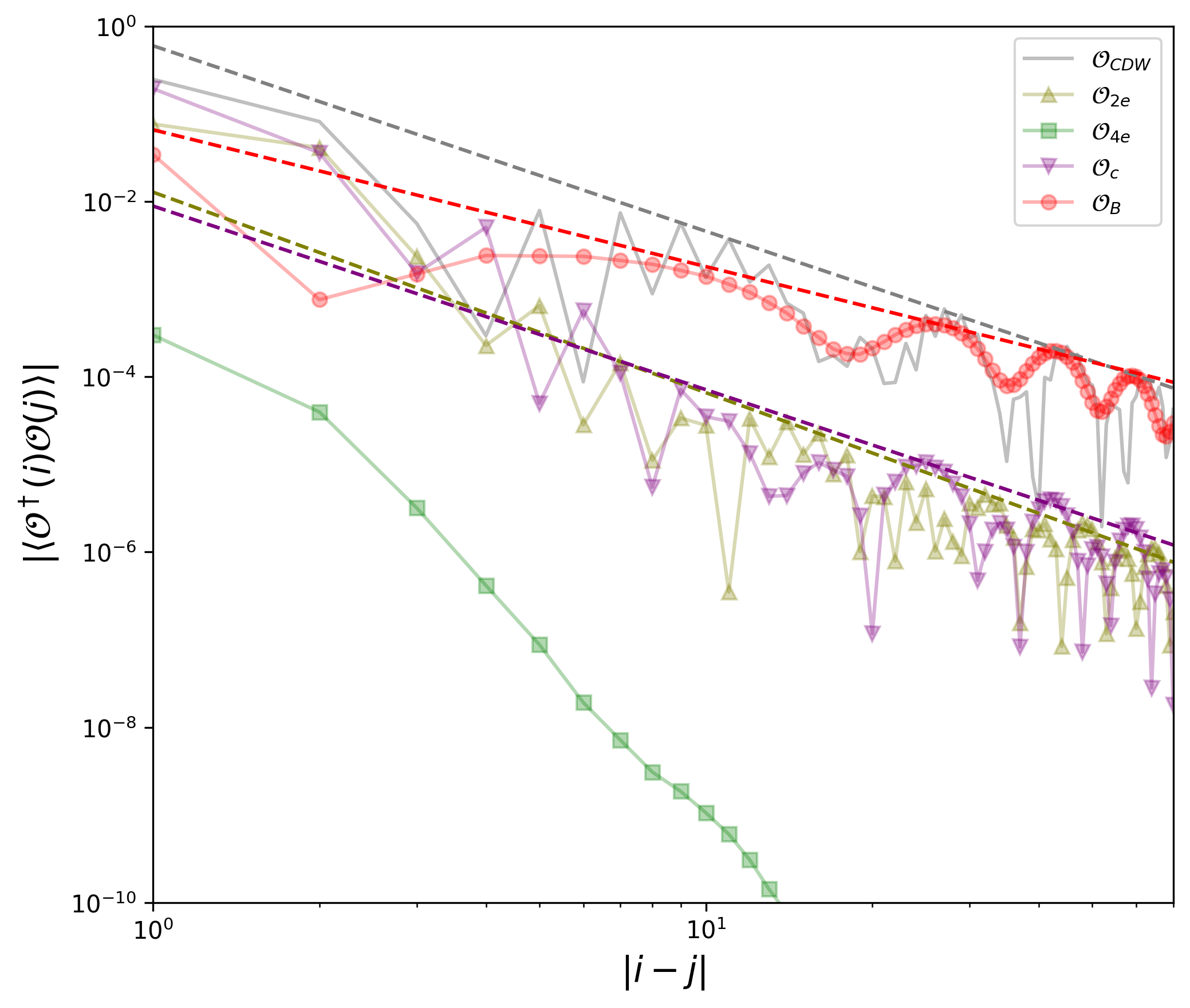}
\caption{Correlation functions $|\braket{\mathcal{O}^{\dagger}(i)\mathcal{O}(j)}|$ for (top) $S=1/2$ and (bottom) $S=3/2$. The dashed lines are fits to the calculated curves.}
\label{fig:correlation_infinite}
\end{figure}
\begin{figure}[htb]
\centering
    \includegraphics[width=0.8\columnwidth]{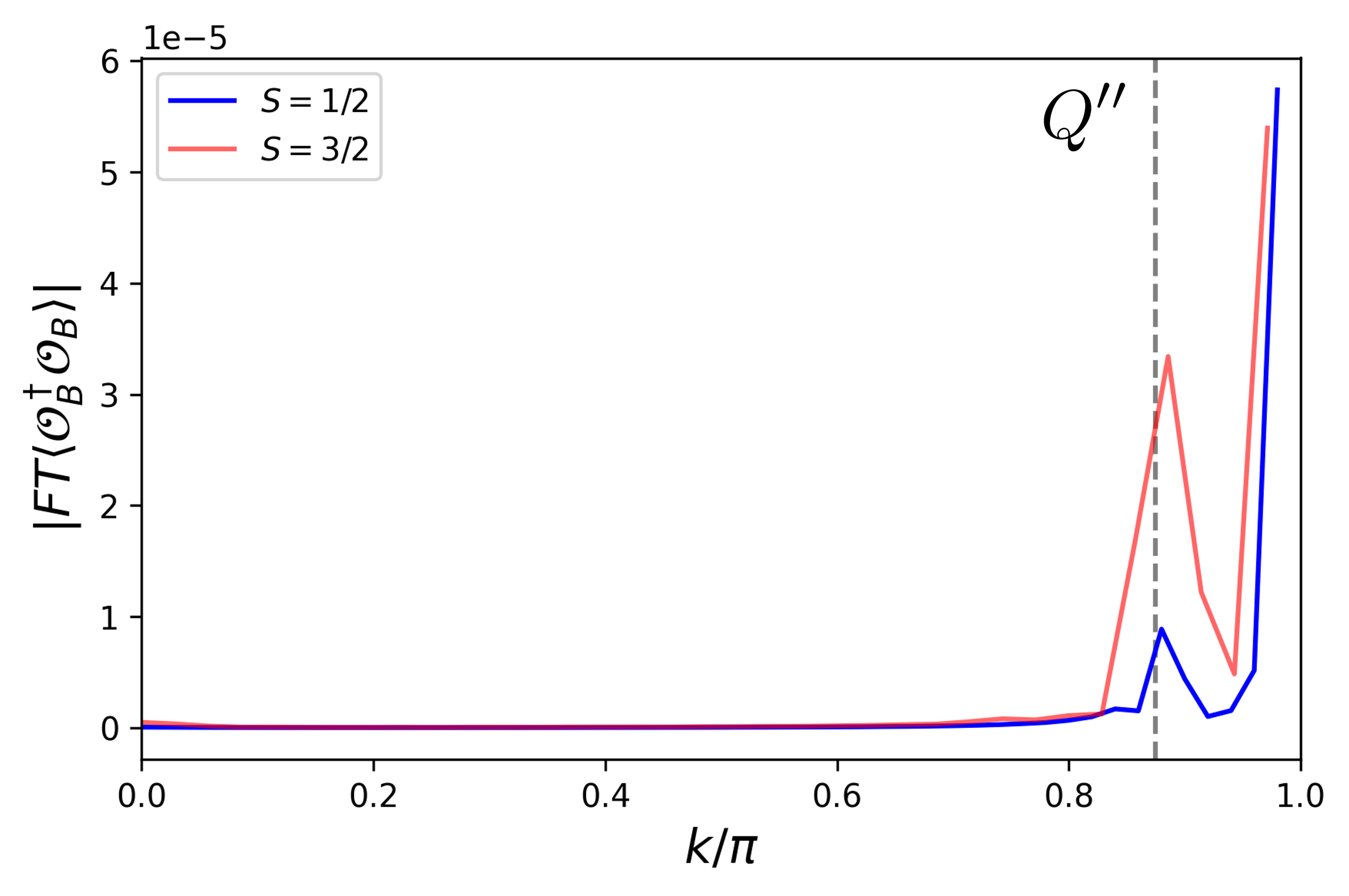}
\caption{Fourier spectra of the bond-pairing correlation functions for $S=1/2$ and $S=3/2$. There are peaks at $Q=\pi$ and $Q''=2k_F$.}
\label{fig:FT_pairing}
\end{figure}

Figure~\ref{fig:correlation_infinite} shows the calculated correlation functions $\braket{\mathcal{O}^{\dagger}(i)\mathcal{O}(j)}$.
For both the $S=1/2$ and $S=3/2$ systems, we find that the bond-pairing exhibits the power-law behavior $\braket{\mathcal{O}_{\rm B}^{\dagger}(i)\mathcal{O}_{\rm B}(j)}\sim (\mbox{oscillation})\times |i-j|^{-\alpha_{\rm B}}$ with the smallest exponents among the operators considered, $\alpha_{\rm B}\simeq 1.11$ for $S=1/2$ and $\alpha_{\rm B}\simeq 1.56$ for $S=3/2$.  This means that it is indeed the most dominant quasi-long-range order in agreement with the previous studies for finite-size systems under the open boundary conditions~\cite{Berg2010,Fradkin2014,Fradkin2020}. The bond-pairing correlation functions show spatial oscillation with the wavevectors $Q=\pi$ and $Q''=2k_F=7\pi/8=0.875\pi$ as seen in Fig.~\ref{fig:FT_pairing}, where the Fourier transform is performed using the calculation data up to distance $L_{\rm FT}=100$ for $S=1/2$ and $L_{\rm FT}=70$ for $S=3/2$. The oscillating behaviors and the absence of a uniform component mean that the bond-pairing state is indeed a PDW state. We stress that the dominance of the PDW order is not a trivial result, and the dominant order can be identified clearly with iDMRG, whereas this is difficult in finite DMRG, as will be discussed in the next section (Sec.~\ref{sec:boundary}).
At the present parameter, the second dominant order is the CDW order in the $S=1/2$ system and its exponent is $\alpha_{\rm CDW}\simeq1.51$ which is rather close to $\alpha_{\rm B}\simeq1.11$. For $S=3/2$, the PDW order with $\alpha_{\rm B}\simeq 1.56$ is clearly dominant over the CDW order with $\alpha_{\rm CDW}\simeq 2.12$. The power-law behaviors of the CDW correlation functions at long distance mean that the system preserves the translation symmetry and the CDW is not a true long-range order, which is clearly seen in the present iDMRG calculations but is obscure in a finite-size system (see Sec.~\ref{sec:boundary}).
The composite-pairing correlation has the exponents $\alpha_{c}\simeq2.07$ for $S=1/2$ and $\alpha_{c}\simeq2.10$ for $S=3/2$. 
We note that the composite and charge-$2e$ pairing correlations also exhibit oscillating behaviors at long distance, and therefore they could be interpreted as PDWs. However, it turns out that Fourier components of these pairing correlations do not have clear peak structures and instead have broad weights over the whole Brillouin zone.
For the bond-pairing, it was proposed that the $\pi$-PDW is a composite order of the $\eta$-pairing with the momentum $-2k_F$ and a (charge-0, spin-0) excitation with the momentum $P_{\rm YOA}=2k_F+\pi$~\cite{Berg2010}. Although this is a reasonable argument in the weak Kondo coupling region $J_{K}\ll J_H$, a small non-zero $J_K$ could be potentially dangerous and change the Fermi surface structure as in the pure Kondo lattice without the Heisenberg interaction~\cite{Tsunetsugu1997}. Since the two coupling constants are comparable in the present study, $J_K=J_H=2t$, the Fermi surface structure may be non-trivial. We will come back to this point soon.

\begin{figure}[htb]
    \centering
    \includegraphics[width=0.8\linewidth]{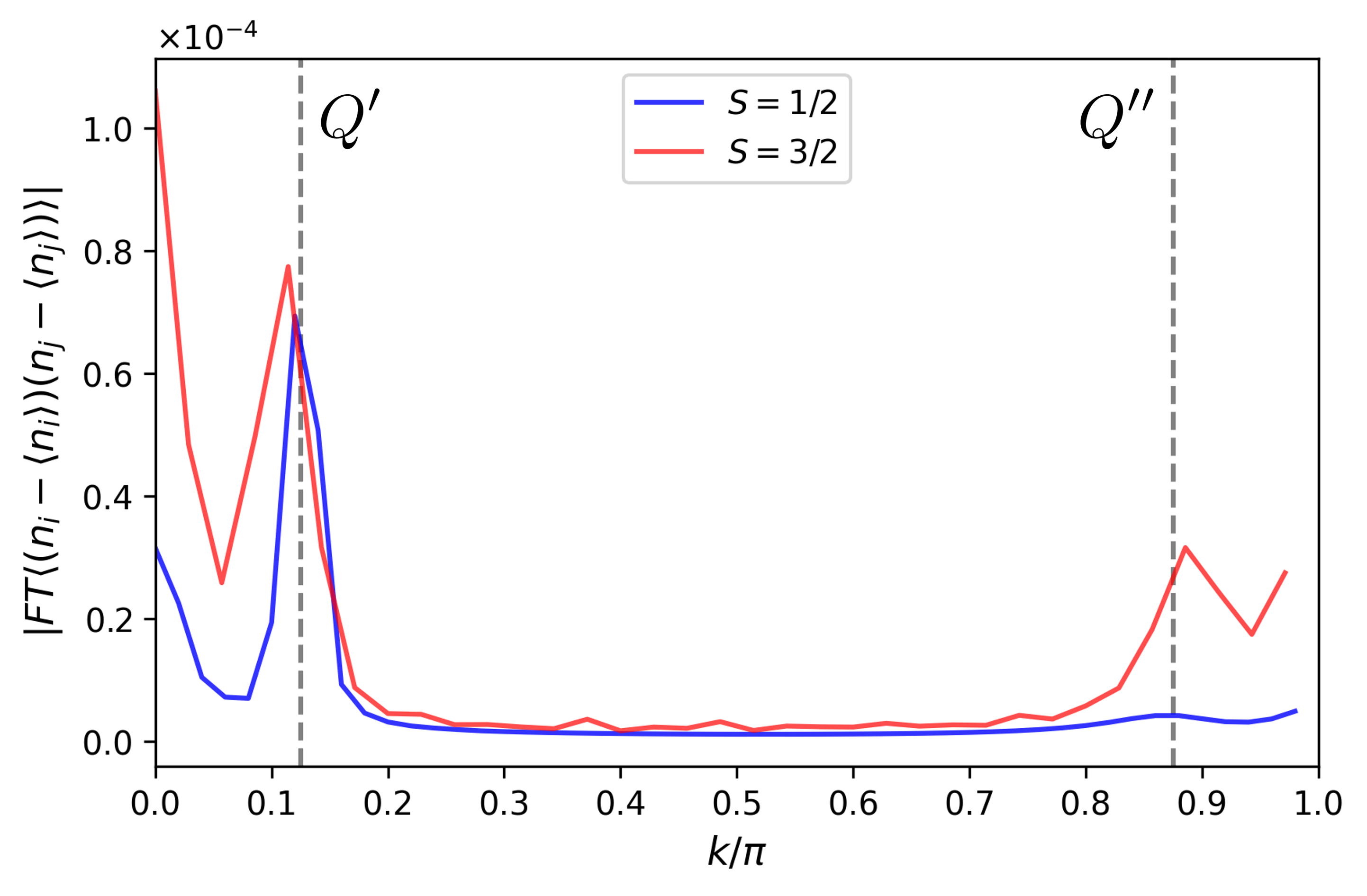}
    \caption{Fourier spectra of the density-density correlation function. The wavevectors corresponding to the peaks are $Q=\pi, Q'=-(2k_F+\pi)= 0.125\pi$, and $Q''=2k_F=0.875\pi$ modulo $2\pi$. The peak at $k=0$ corresponds to the uniform mode. }
  \label{fig:FT_density_func}
\end{figure}

The CDW correlation function exhibits oscillations with the wavevectors $Q=\pi, Q'=-(2k_F+\pi)$, and $Q''=2k_F$ as seen in the Fourier spectra (Fig.~\ref{fig:FT_density_func}). The peaks at $Q$ and $Q''$ are small for the $S=1/2$ system, but they are clearly visible for the $S=3/2$ system.
These three wavevectors satisfy the relation $Q+Q'+Q''=0$. This suggests that the three gapless density modes are not independent and are correlated in a non-trivial way. One may expect that the two modes at $Q=\pi$ and $Q''=2k_F$ belong to the same class since their amplitudes are comparable, while the mode at $Q'=-(2k_F+\pi)$ has a stronger amplitude. Indeed, according to the YOA argument, there exists a gapless excitation with the momentum $P_{\rm YOA}=2k_F+\pi$ carrying charge-0 and spin-0~\cite{Yamanaka1997}. 
The strongest peak occurs at \(Q'=-(2k_F+\pi)\), which is precisely the YOA momentum modulo \(2\pi\). This indicates that the YOA-required scalar mode has a visible overlap with the density channel. This picture is further supported from the viewpoint of the Fermi surface structure as discussed below.

\begin{figure}[htb]
    \centering
    \includegraphics[width=0.8\linewidth]{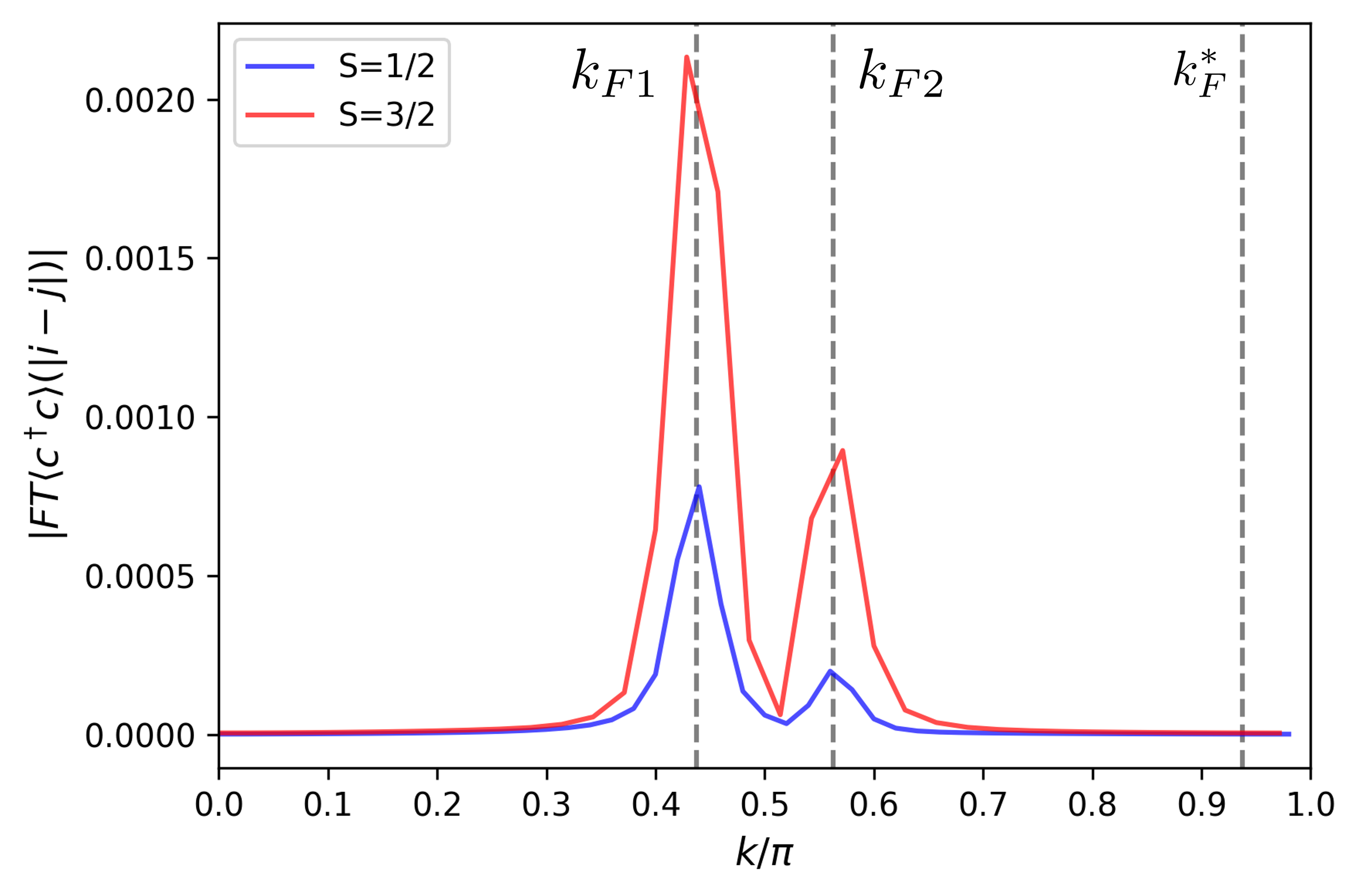}
    \caption{Fourier spectra of the single-particle correlation function. The wavevectors corresponding to the peaks are $k_{F1}=k_F\simeq 0.43\pi$, and $k_{F2}=Q-k_F\simeq 0.57\pi$. There is no peak at $k_F^*=k_F+\pi/2\simeq 0.94\pi$. }
  \label{fig:FT_single_particle}
\end{figure}
\begin{figure}[htb]
    \centering
    \includegraphics[width=0.8\linewidth]{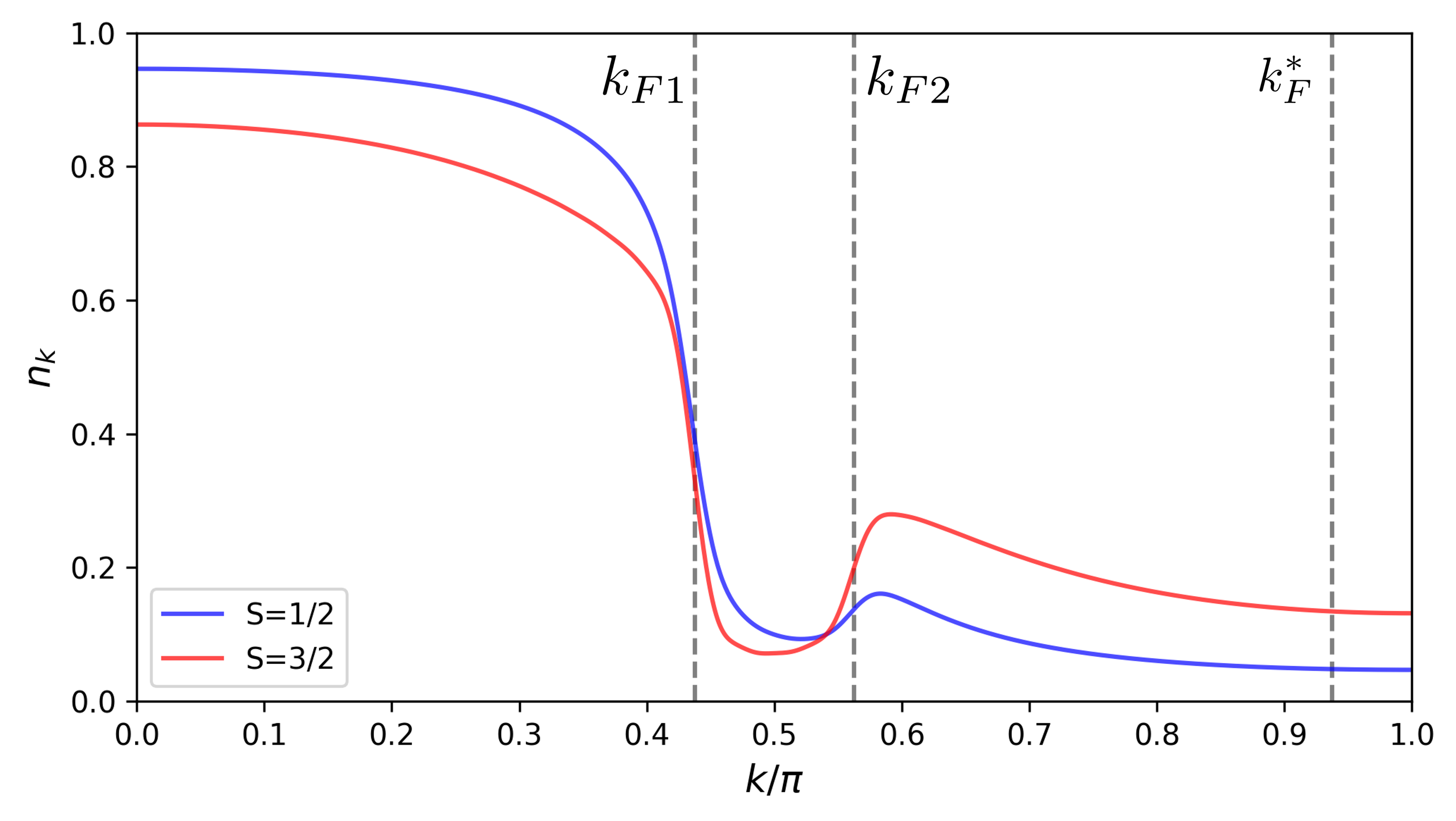}
    \caption{Momentum distribution function. The dashed vertical lines indicate the momenta $k_{F1}=k_F\simeq 0.44\pi, k_{F2}=Q-k_F\simeq 0.56\pi$, and $k_F^*=k_F+\pi/2\simeq0.94\pi$, respectively.}
  \label{fig:nk_PDW}
\end{figure}

We investigate the single-particle correlation function $G(i-j)=\braket{c_{i\alpha}^{\dagger}c_{j\alpha}}$ and show its Fourier spectrum in Fig.~\ref{fig:FT_single_particle}. Clearly, there are two peaks at $k_{F1}=k_F\simeq0.44\pi$ and $k_{F2}=Q-k_F\simeq 0.56\pi$, while no peak is found at $k_F^*=k_F+\pi/2\simeq 0.94\pi$. We stress that the single-particle correlation function contains two power-law decaying components,
\begin{align} 
G(x)\simeq A_{1}\frac{\cos(k_{F1}x+\varphi_1)}{x^{\alpha_1}}+A_{2}\frac{\cos(k_{F2}x+\varphi_2)}{x^{\alpha_2}},
\label{eq:Gx}
\end{align}
where $A_{1,2}$, $\varphi_{1,2}$, and $\alpha_{1,2}$ are amplitudes, phase shifts, and exponents, respectively.
This means that the single-particle correlation contains contributions from two gapless fermionic components at the distinct momenta \(k_{F1}\) and \(k_{F2}\).
This is in stark contrast to the Kondo-decoupled limit $J_K=0$ where there is a single fermionic mode at $k_F$. Numerically, the exponents are evaluated as $(\alpha_1,\alpha_2)\simeq(1.87, 2.26)$ for $S=1/2$ and $(\alpha_1,\alpha_2)\simeq(1.7, 2.0)$ for $S=3/2$.

The underlying Fermi surface structure can be understood based on the momentum distribution.
Figure~\ref{fig:nk_PDW} shows the single-particle momentum distribution function 
\begin{align}
n(k)=\frac{1}{L_{\rm FT}}\sum_{ij,\alpha}\braket{c^{\dagger}_{i\alpha}c_{j\alpha}}e^{ik(i-j)}.
\end{align} 
It is non-monotonic as a function of the momentum $k$ in sharp contrast to non-interacting fermions and also conventional interacting fermions. 
Qualitatively similar behaviors  have been obtained in the previous studies for the finite-size systems~\cite{Moukouri1996,Edelstein2011,Yang2025}. For the $S=1/2$ system, it seems that there is a hump-like structure around $k_{F2}$. However, for $S=3/2$, the structure is sharper and one can clearly see that there is a dip in $n(k)$ around $k=\pi/2$. Quantitatively, the difference $\Delta n=n(k=\pi/2)-n(k=\pi)$ can be a useful indicator of dip/hump structures. We find $\Delta n=0.053>0$ for $S=1/2$ while $\Delta n=-0.060<0$ for $S=3/2$. The clear dip structure provides clear evidence for an interior-gap-like reconstruction of the single-particle momentum distribution. This strongly suggests that the PDW state at the present parameters is accompanied by an interior-gap-like single-particle structure. This is in sharp contrast to the previous studies of the interior-gap superconductivity where it has been considered that an interior-gap state and a non-uniform state such as a Fulde-Ferrell-Larkin-Ovchinnikov state are competing and do not coexist~\cite{Sarma1963,Liu2003,Wu2003,Bedaque2003,Gubankova2003,Forbes2005,Gubbels2013,Liu2022,FF1964,LO1965}.
We note that a PDW state does not automatically imply an interior-gap structure. Finite-momentum pairing generally reconstructs the single-particle spectrum, but whether gapless pockets remain is model dependent~\cite{Baruch2008,Berg2009prb,Fradkin2015,Agterberg2020}. The present result is therefore nontrivial in that the Kondo-Heisenberg-chain PDW is shown numerically to realize such an interior-gap-like reconstruction.

Furthermore, the dip structure of $n(k)$ and the Fourier spectrum of $G(x)$ together imply that there are two Fermi surfaces at $k_{F1}$ and $k_{F2}$, while there is no Fermi surface at $k_F^*$. This behavior was also pointed out based on the single-particle spectral function in the previous study~\cite{Yang2025} and is consistent with the Cooper-pair picture between $k_F$ and $Q-k_F$ for non-uniform superconductivity. The existence of two small Fermi surfaces and the absence of a large Fermi surface are highly unusual. In the standard half-odd-integer $S$ Kondo lattice at $J_H=0$, it is well-known that there is a single large Fermi surface with the Fermi wavevector $k_F^*=k_F+\pi/2$~\cite{Tsunetsugu1997}.
It is important to distinguish the YOA momentum constraint from the existence of a single-particle large Fermi surface. The YOA argument guarantees a low-energy charge-neutral and spin-neutral excitation at momentum
\begin{align}
P_{\rm YOA}=2k_F+\pi=2k_F^\ast
\end{align}
for the half-odd-integer-spin Kondo-Heisenberg chain~\cite{Yamanaka1997}. However, this statement alone does not require a singularity in the single-particle Green's function at \(k_F^\ast\). The latter identification is an additional Fermi-surface interpretation of the YOA momentum, which is natural in a conventional large-Fermi-surface state but is not a logical consequence of the theorem itself. Our numerical results provide a concrete example of this distinction: the momentum distribution and the single-particle correlation function show no singular structure at \(k_F^\ast\), while the density correlations clearly contain the YOA wavevector. Thus, in the present spin-gapped PDW phase, the YOA-required excitation is realized as a composite scalar mode rather than as a conventional single-particle large Fermi surface.

In the present system, the two Fermi surfaces can lead to density excitations with $2k_{F1}=2k_F, 2k_{F2}=2\pi-2k_F=-2k_F$, and also $k_{F1}+k_{F2}=\pi$. The former two excitations are partner modes and each of them is generated as a particle-hole excitation around a Fermi surface. On the other hand,  the latter mode is an inter-Fermi-surface particle-hole excitation connecting the two Fermi surfaces. A composite mode of $2k_F$- and $\pi$- modes with the momentum $2k_F+\pi$ can exist as a part of the YOA excitation. 
Similarly, as already mentioned, 
a natural interpretation of the present numerical results is that the PDW state is formed by pairing between electrons on the two Fermi surfaces at \(k_{F1}=k_F\) and \(k_{F2}=Q-k_F\)~\cite{Yang2025}. In this sense, the $\pi$-PDW can be viewed as a \((k,Q-k)\)-pairing state, which is the standard momentum-space picture of nonuniform superconductivity. At the same time, this interpretation is consistent with the earlier viewpoint, in which the \(\pi\)-PDW is understood as a composite of the \(\eta\)-pairing channel and the YOA momentum~\cite{Berg2010}.

\begin{figure}[htb]
\centering
    \includegraphics[width=0.8\columnwidth]{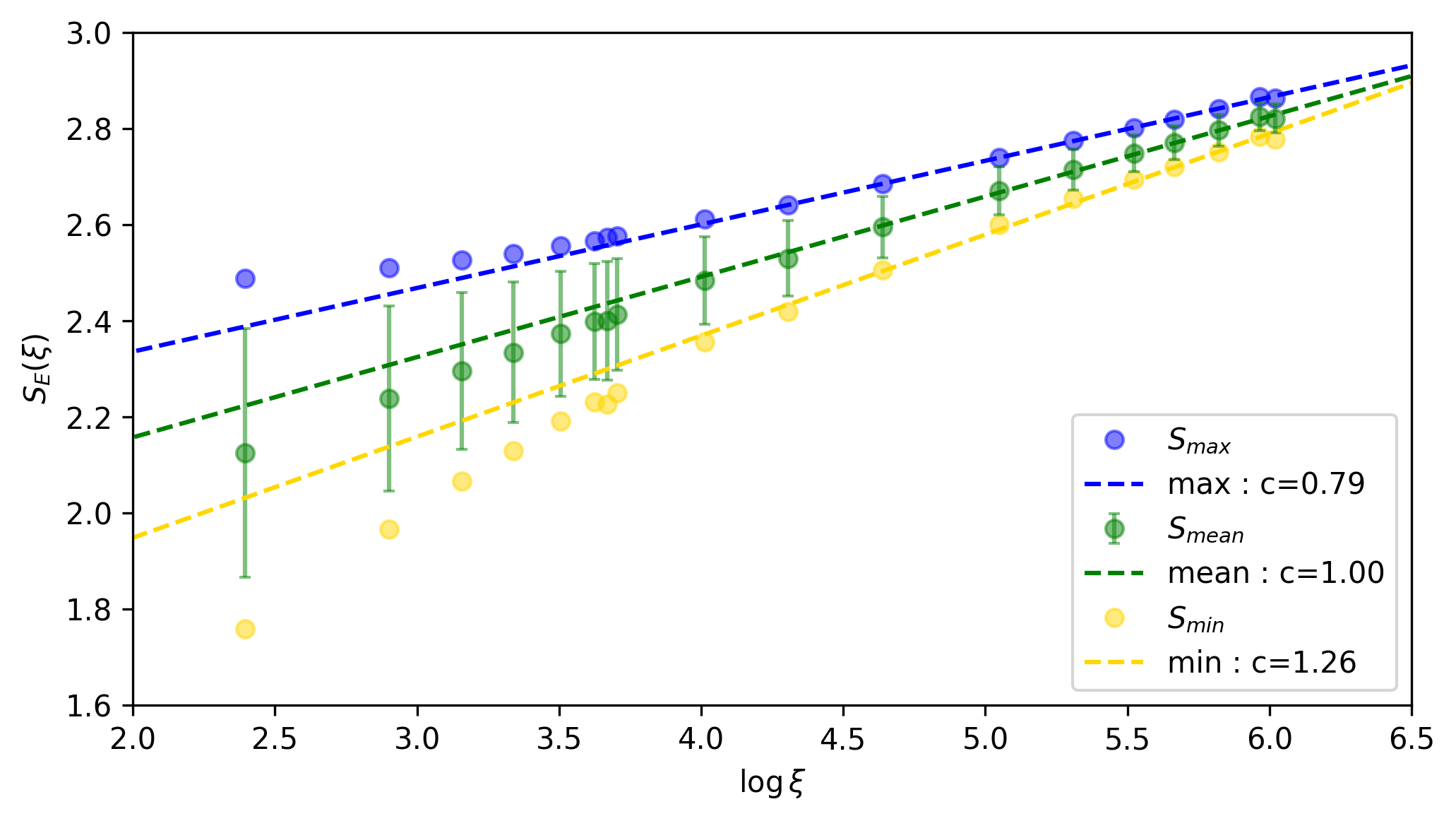}
    \includegraphics[width=0.8\columnwidth]{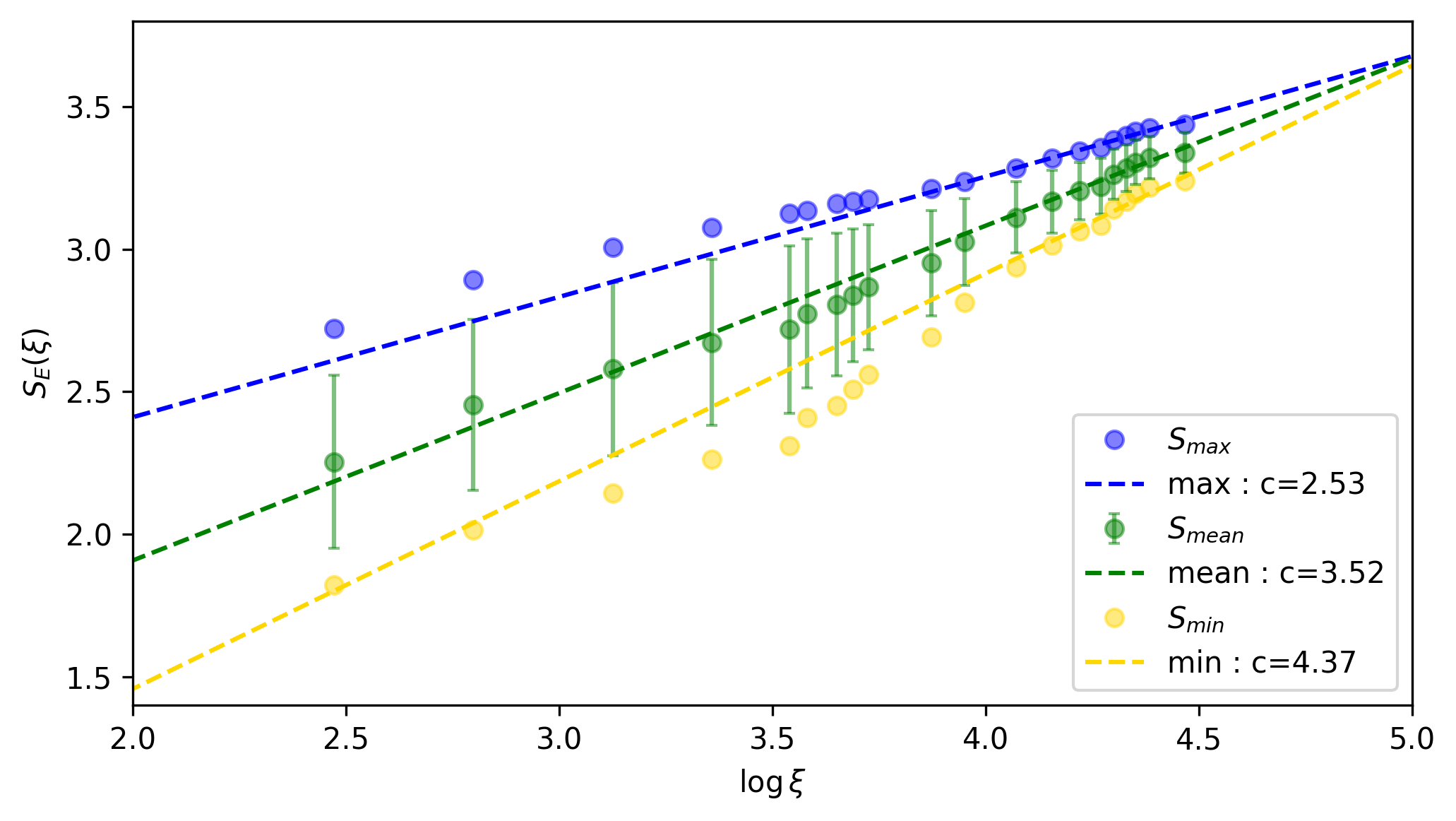}
\caption{Entanglement entropy $S_E$ for (top) $S=1/2$ and (bottom) $S=3/2$ Kondo-Heisenberg model. The fitting regions are $\xi > 60$ ($\log \xi > 4.0$). The largest bond dimensions are $\chi=9000$ for $S=1/2$ and $\chi=11000$ for $S=3/2$.}
\label{fig:EE}
\end{figure}

To further develop our understanding of the Fermi surface structure and low energy excitations, we evaluate the central charge $c$ based on scaling of the entanglement entropy with respect to the correlation length $\xi$ of the transfer matrix~\cite{CalabreseCardy2004},
\begin{align}  
S_{E}=\frac{c}{6}\log \xi +S_0,
\end{align}
where $S_0$ is a non-universal constant.
In the present system with the filling $n=7/8$, we use the super unit cell size $\ell=16$ in the iDMRG calculations. The entanglement entropy $S_E(x)$ depends on the bond position $x$ within the super unit cell, $x=1,2,\cdots,16$, especially when the bond dimension is small. Nevertheless, the central charge $c$ should be uniquely determined if the correlation length $\xi$ is sufficiently long and the system is well within the scaling regime, while the non-universal constant $S_0(x)$ can depend on the bond position. We show the numerical results for the maximum $(S_{\rm max})$,  minimum $(S_{\rm min})$, and mean value $(S_{\rm mean})$ of the entanglement entropy within the super unit cell in Fig.~\ref{fig:EE}. For $S=1/2$, a weighted linear fit of the mean value $S_{\rm mean}$ gives the central charge $c\simeq1.00.$ Besides, $S_{\rm max}$ and $S_{\rm min}$ asymptotically have similar slopes for large $\xi$, which also leads to a rough estimate $c\simeq 1.0\pm 0.3$. On the other hand, for $S=3/2$, a fit to $S_{\rm mean}$ gives an unrealistically large value $c\simeq 3.52$, and the difference of the slopes for $S_{\rm max}$ and $S_{\rm min}$ is significant even for the largest $\xi$. Therefore, the data may not yet be in a reliable scaling regime and it is difficult to correctly evaluate the central charge for the $S=3/2$ system. As will be discussed in the next section, the entanglement entropy for the $S=1/2$ model under the open boundary condition gives a consistent result $c\simeq 1.00$, while it does not provide a reliable evaluation for $S=3/2$ due to strong boundary effects.

The numerical result $c\simeq 1.00$ for $S=1/2$ suggests that the central charge is $c=1$ in this system. Thus, despite the presence of the two Fermi surfaces, there is a single U(1) bosonic degree of freedom $(\phi_c,\theta_c)$ for the charge sector from the viewpoint of bosonization. Note that there is no such gapless boson for the gapped spin sector. The low energy physics should be effectively described by a renormalized free boson theory of the dual boson fields $(\phi_c,\theta_c)$. This implies that both of the two fermionic excitations around $k_{F1}$ and $k_{F2}$ are related to $(\phi_c,\theta_c)$ and are not independent at low energy. Furthermore, any gapless modes such as the three oscillation modes in the density correlation function (Fig.~\ref{fig:FT_density_func}) should also be described by $(\phi_c,\theta_c)$. This will hold even for a composite (charge-0, spin-0) mode containing spin operators.

The identification of $c=1$ is not a trivial consequence of the spin gap or of the $c$-theorem~\cite{Zamolodchikov1986,Cardy1996}. While these considerations constrain the infrared theory, they do not by themselves exclude the possibility that the charge sector contains more than one gapless bosonic mode.
In particular, once two Fermi-surface singularities are observed in the single-particle sector, a naive expectation of $c=2$ can arise. The numerical result $c=1$ therefore provides a useful constraint on the low-energy interpretation of the two Fermi-surface singularities.

A useful phenomenological picture suggested by our numerical results is that the PDW order, the emergent Fermi-surface singularity at $k_{F2}$, and the charge-neutral mode at $Q=\pi$ are not independent structures but mutually related aspects of the same correlated phase. In this view, the singularity at $k_{F2}$ is not regarded as a preexisting bare Fermi surface on an equal footing with the original $k_F$ one. Rather, it is naturally interpreted as an emergent composite fermionic component involving the original $k_F$ fermion and a neutral excitation carrying momentum $Q=\pi$. Formally, the fermionic field operator around the $k_{F2}$-Fermi surface is expressed as $\psi_{k_{F2}}\simeq \psi_{-k_{F1}}\Phi_Q$, where $\psi_{-k_{F1}}$ is the left-moving fermionic field around the $k_{F1}$-Fermi surface and $\Phi_Q$ is a gapless bosonic field with momentum $Q=\pi$. Then, the fermion field may be expanded as 
\begin{align}
\psi(x)=e^{ik_{F1}x}\psi_{k_{F1}}(x)+e^{ik_{F2}}\psi_{k_{F2}}(x)+\cdots 
\end{align}
at low energy.
As mentioned above, the numerically obtained central charge $c=1$ implies that all the low energy degrees of freedom should be written in terms of the boson field $(\phi_c,\theta_c)$. We may naively regard that the $\psi_{k_{F1}}$-field corresponds to the preexisting fermionic excitation around the small Fermi surface at $J_K=0$. In this picture, $\psi_{k_{F1}}$-field is the fundamental field and $\psi_{k_{F2}}$-field describes emergent fermions, which is consistent with the numerical results that the exponents of $G(x)$ (Eq.~\eqref{eq:Gx}) satisfy $\alpha_1<\alpha_2$. Given the spin gap, a natural candidate for the neutral excitation $\Phi_Q$ is a scalar mode in the charge sector. At the same time, the same $Q=\pi$ structure also appears in the dominant pairing channel as the PDW wavevector. From this perspective, the PDW order, the reconstructed single-particle structure, and the $Q=\pi$ neutral mode should be viewed as intertwined manifestations of one and the same strong-coupling state, rather than as a simple cause-and-effect sequence among separate ingredients.

To clarify the implication of the YOA constraint, it is useful to contrast this situation with the notion of a fractionalized Fermi liquid (FL$^\ast$)~\cite{Senthil2003,Vojta2018}. In a higher-dimensional FL$^\ast$, the Oshikawa-type momentum balance can be satisfied without a conventional large Fermi surface~\cite{Oshikawa2000}, because part of the momentum is carried by a topological or fractionalized sector rather than by ordinary quasiparticles. The present one-dimensional state is different: the additional momentum component appears as an actual gapless neutral mode at \(Q=\pi\), which is visible in the density correlations and is intertwined with both the PDW correlation and the emergent \(k_{F2}\) structure. 
Thus, the comparison emphasizes that the YOA momentum constraint can be satisfied without being realized as a conventional single-particle large Fermi surface.

Although the above phenomenological understanding is consistent with our numerical results, one could also perform bosonization procedure based on the four Fermi points found in the previous study~\cite{Yang2025}. In this case, each pair of the left- and right-moving fermions for the two Fermi surfaces would behave as low energy independent degrees of freedom, and consequently the central charge could be $c=2$. Therefore, a naive application of the standard four-Fermi-point bosonization scheme would be insufficient for the present interior-gap PDW state with the emergent Fermi surface. A complete formulation of a bosonization theory will be a challenging and interesting future study.

Finally, we note that the term ``interior-gap'' in the present system should not be understood in the literal geometric sense familiar in higher dimensions, where a partially gapped Fermi surface leaves gapless pockets~\cite{Sarma1963,Liu2003,Wu2003,Bedaque2003,Gubankova2003,Forbes2005,Gubbels2013,Liu2022}. In one dimension, the Fermi surface reduces to Fermi points, and the corresponding notion is instead a paired state with a reconstructed single-particle spectrum that still retains gapless fermionic singularities.

\subsection{Boundary-induced structures in finite systems}
\label{sec:boundary}

\begin{figure}[htb]
\centering
    \includegraphics[width=0.8\columnwidth]{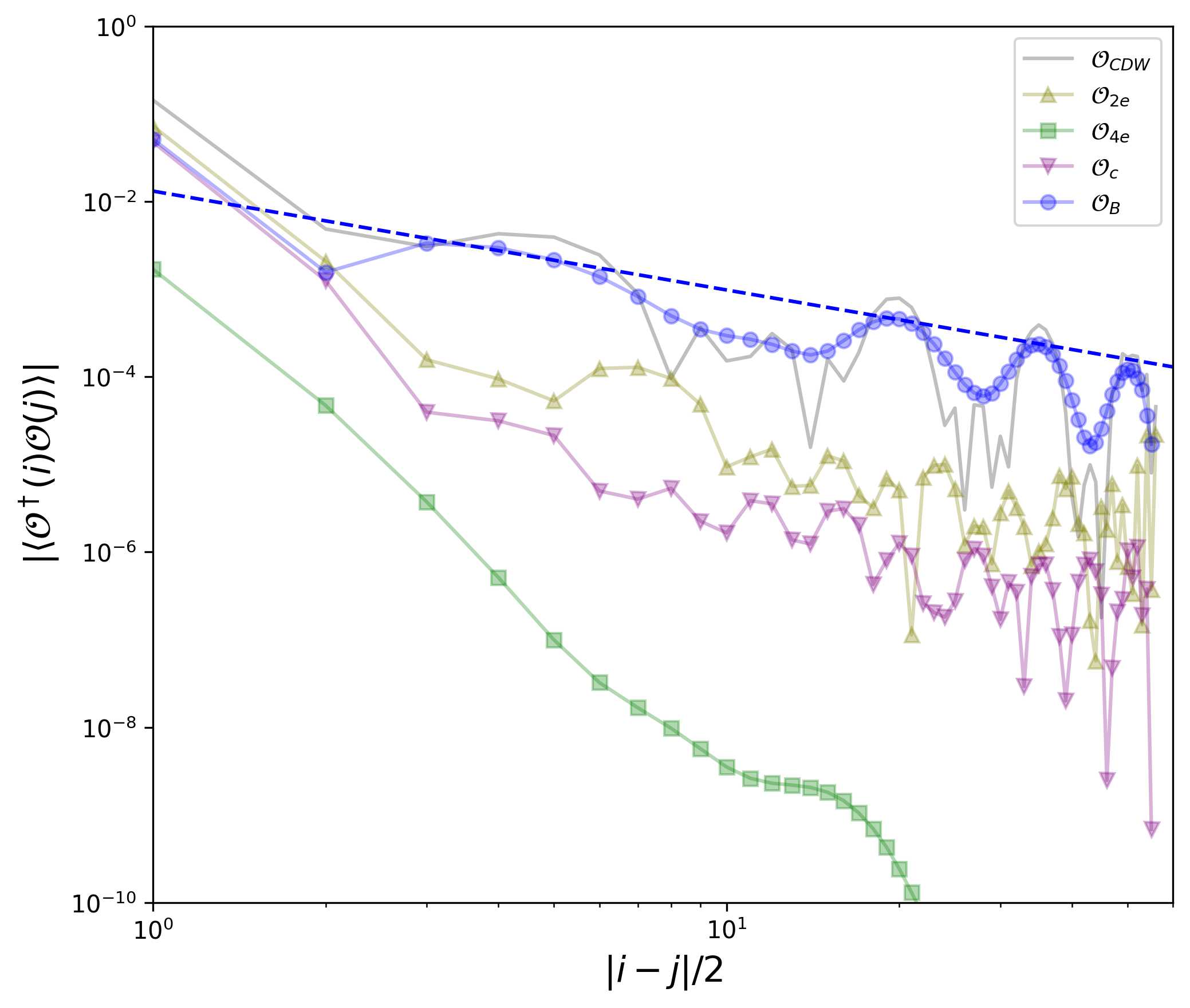}
    \includegraphics[width=0.8\columnwidth]{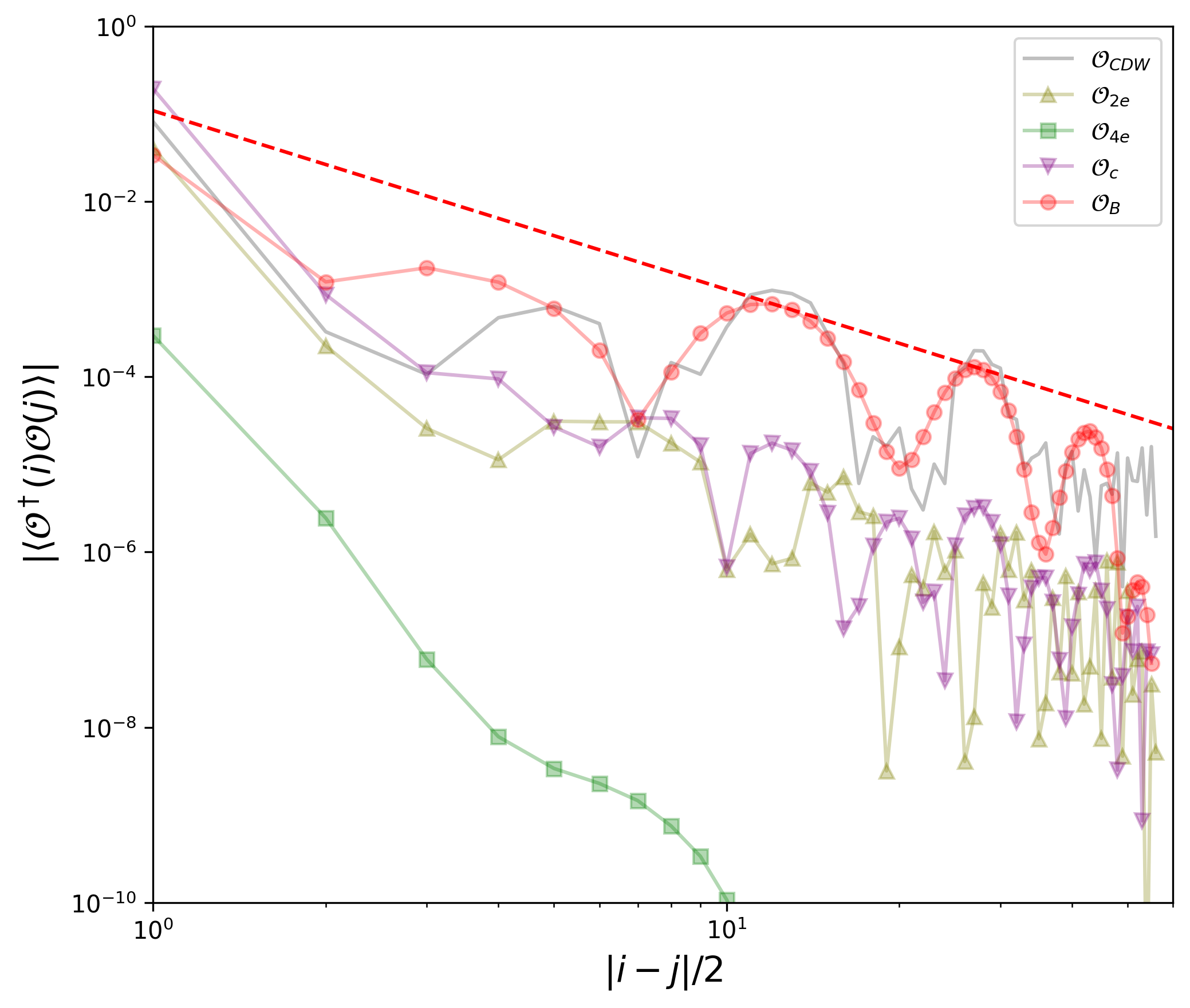}
\caption{Correlation functions $|\braket{\mathcal{O}^{\dagger}(i)\mathcal{O}(j)}|$ for (top) $S=1/2$ and (bottom) $S=3/2$. The dashed line shows a fit to the calculated curve for the PDW correlation function.}
\label{fig:correlation_finite}
\end{figure}

The purpose of this subsection is not to re-identify the bulk phase, but to demonstrate how open boundaries can distort the apparent hierarchy of correlations and affect physical quantities.
In a finite-size system with the open boundary condition, the physical edges can generally induce non-trivial effects. We show the correlation functions $\braket{\mathcal{O}^{\dagger}(i)\mathcal{O}(j)}$ with $i=L/2-x, j=L/2+x$ in Fig.~\ref{fig:correlation_finite}. For both $S=1/2$ and $S=3/2$, the CDW, PDW, composite pairing, and charge-$2e$ pairing correlation functions are competitive and it is difficult to identify the most dominant correlation in contrast to the iDMRG results in the previous section (Fig.~\ref{fig:correlation_infinite}). In addition, the composite and charge-$2e$ correlation function for $S=1/2$ show upturn behaviors around $|i-j|\simeq L$, which makes it more difficult to evaluate the correct power-law exponents. They suggest that there are strong boundary effects in the PDW phase of the Kondo-Heisenberg model. Note that our numerical results are consistent with the previous study where they overlap~\cite{Berg2010}.

\begin{figure}[htb]
    \includegraphics[width=0.8\columnwidth]{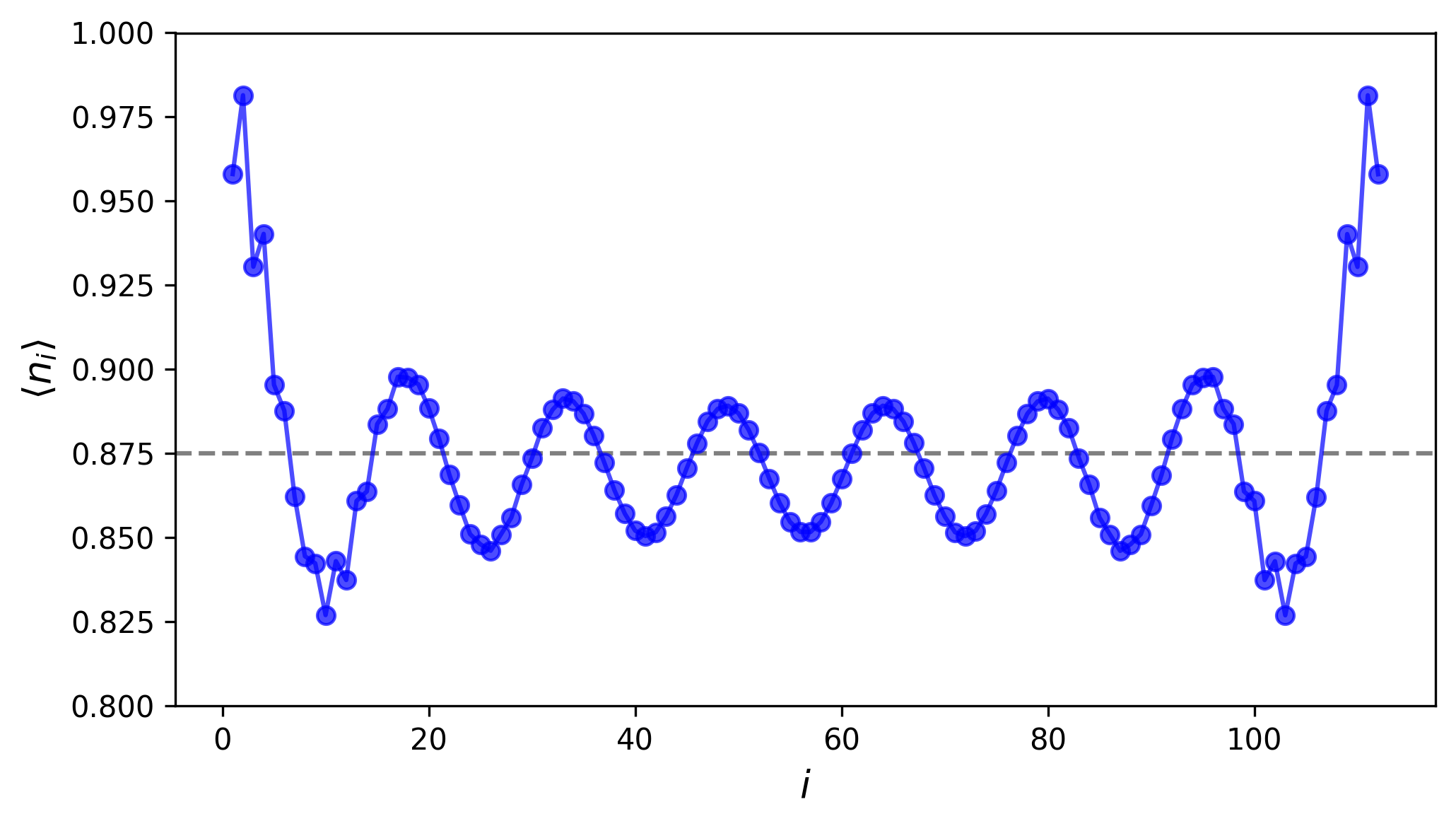}
    \includegraphics[width=0.8\columnwidth]{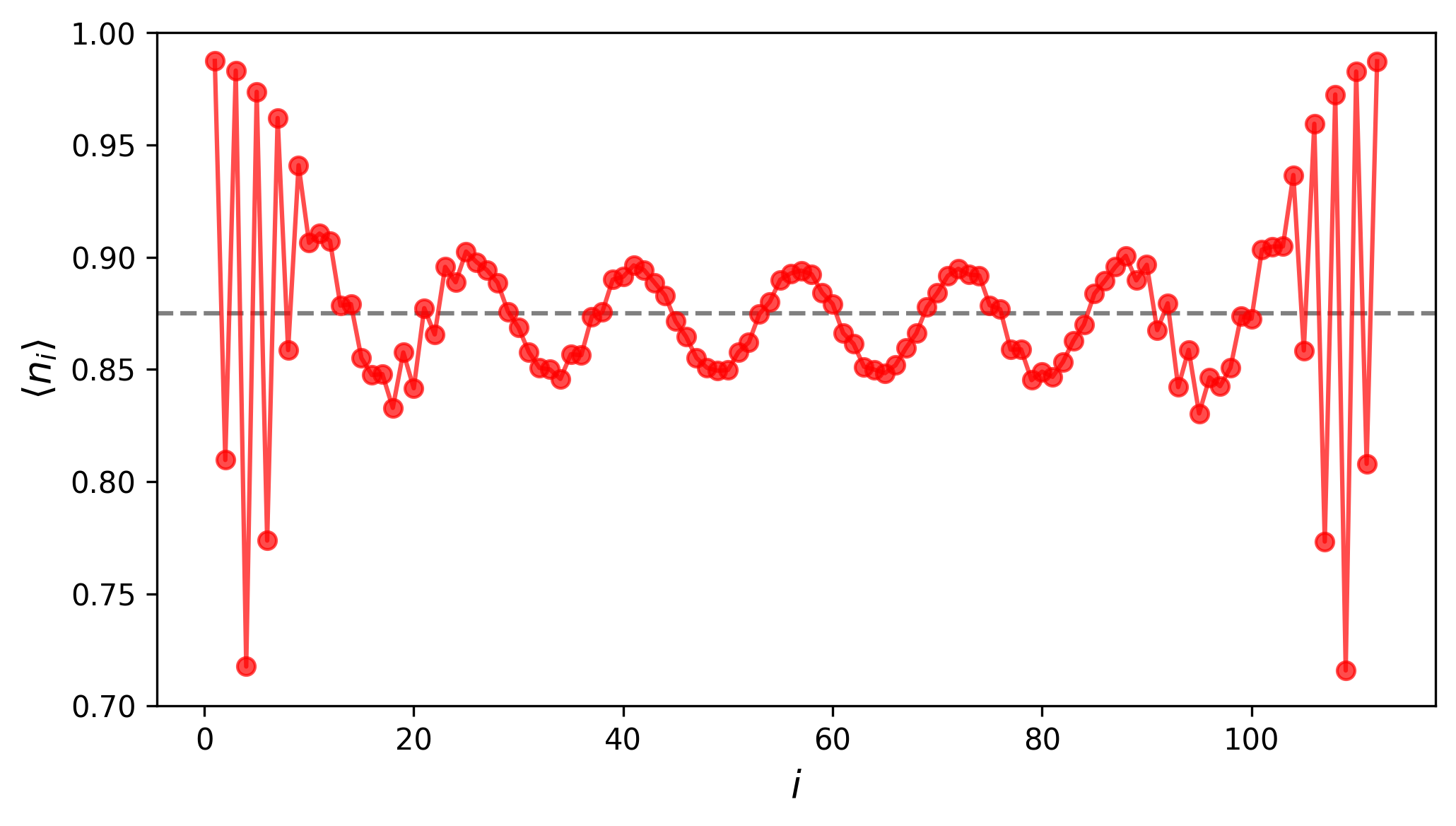}
\caption{Charge density profile $\braket{n_j}$ for (top) $S=1/2$ and (bottom) $S=3/2$.
The dashed horizontal lines indicate the average density $n=7/8$.}
\label{fig:nj}
\end{figure}

By closely looking at the PDW correlation function,
we find that it is suppressed at $|i-j|\simeq L$ in the $S=3/2$ system. The suppression of the PDW correlation is related to the particle density profile in the presence of the boundaries shown in Fig.~\ref{fig:nj}. For $S=1/2$, the particle density $\braket{n_j}$ exhibits a rather standard Friedel oscillation mainly with the wavevector $Q'=2k_F+\pi$ corresponding to the YOA excitation, although there is no large Fermi surface. There are additional oscillations with $Q=\pi$ and $Q''=2k_F$ near the boundaries, but their amplitudes are sufficiently small compared to that of the $Q''$-oscillation. On the other hand, for $S=3/2$, the density profile around the boundary is strongly modified and the $Q,Q''$-oscillations are enhanced. Because the resulting $\braket{n_j}$ exhibits almost a staggered oscillation and the neighboring-site densities differ strongly near the boundary, the inter-site pairing $\mathcal{O}_{\rm B}(j)$ around the edges $j\simeq1$ and $j\simeq L$ is unfavored due to the density mismatch between the sites $j$ and $j+1$. Therefore, the PDW correlation function is suppressed when $i\simeq 1$ and $j\simeq L$, or equivalently $i-j\simeq L$. We suppose that the quantitative difference between $S=1/2$ and $S=3/2$ is non-universal and depends on the amplitudes of the $Q,Q''$-oscillations. These amplitudes are relatively small for $S=1/2$, while they are rather strong for $S=3/2$.
Note that there is no CDW long-range order as already discussed in the previous section, although $\braket{n_j}$ in a finite open chain alone does not unambiguously distinguish a genuine CDW tendency from a boundary-pinned response due to the Friedel oscillations.

\begin{figure}[htb]
\centering
    \includegraphics[width=0.8\columnwidth]{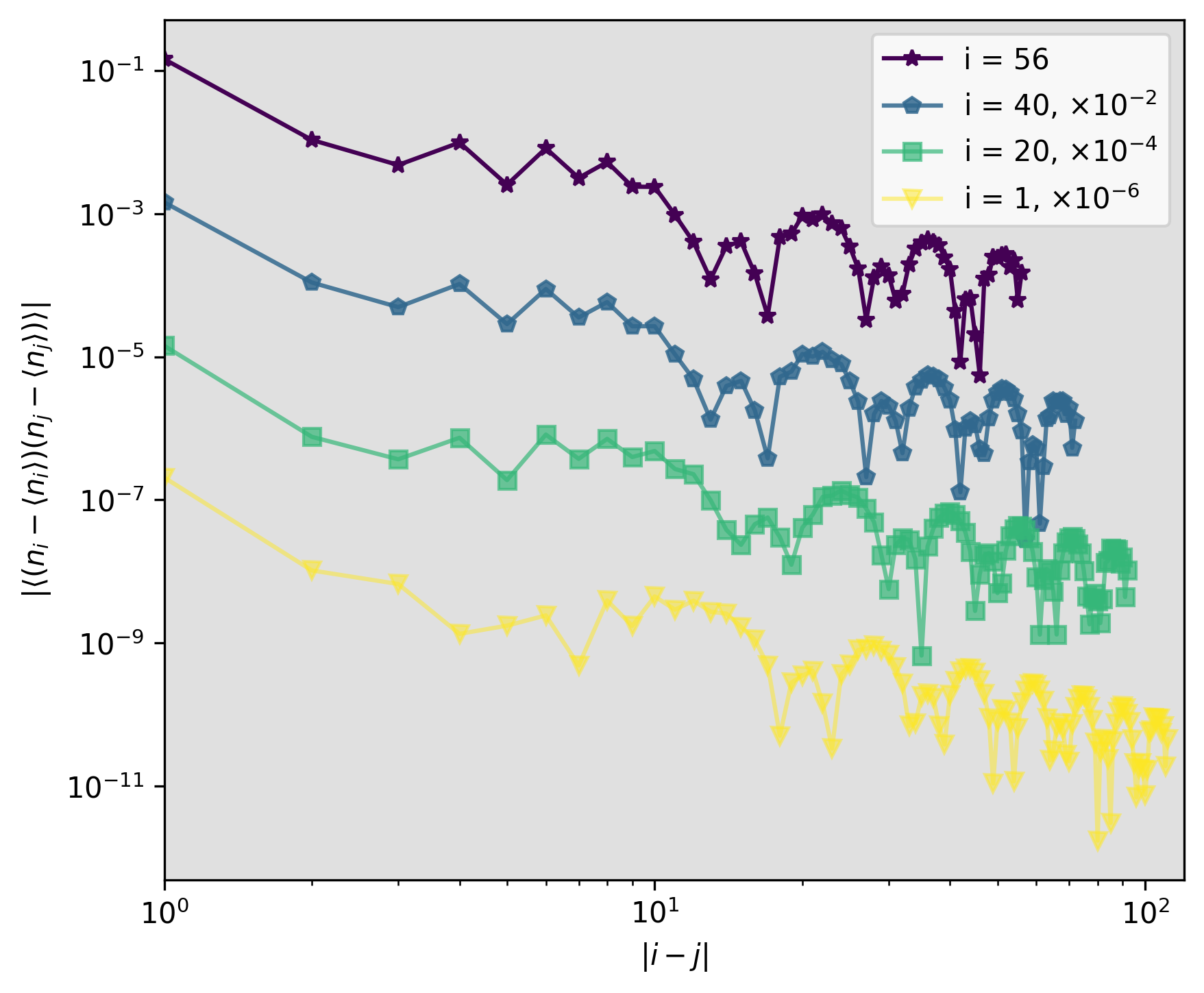}
    \includegraphics[width=0.8\columnwidth]{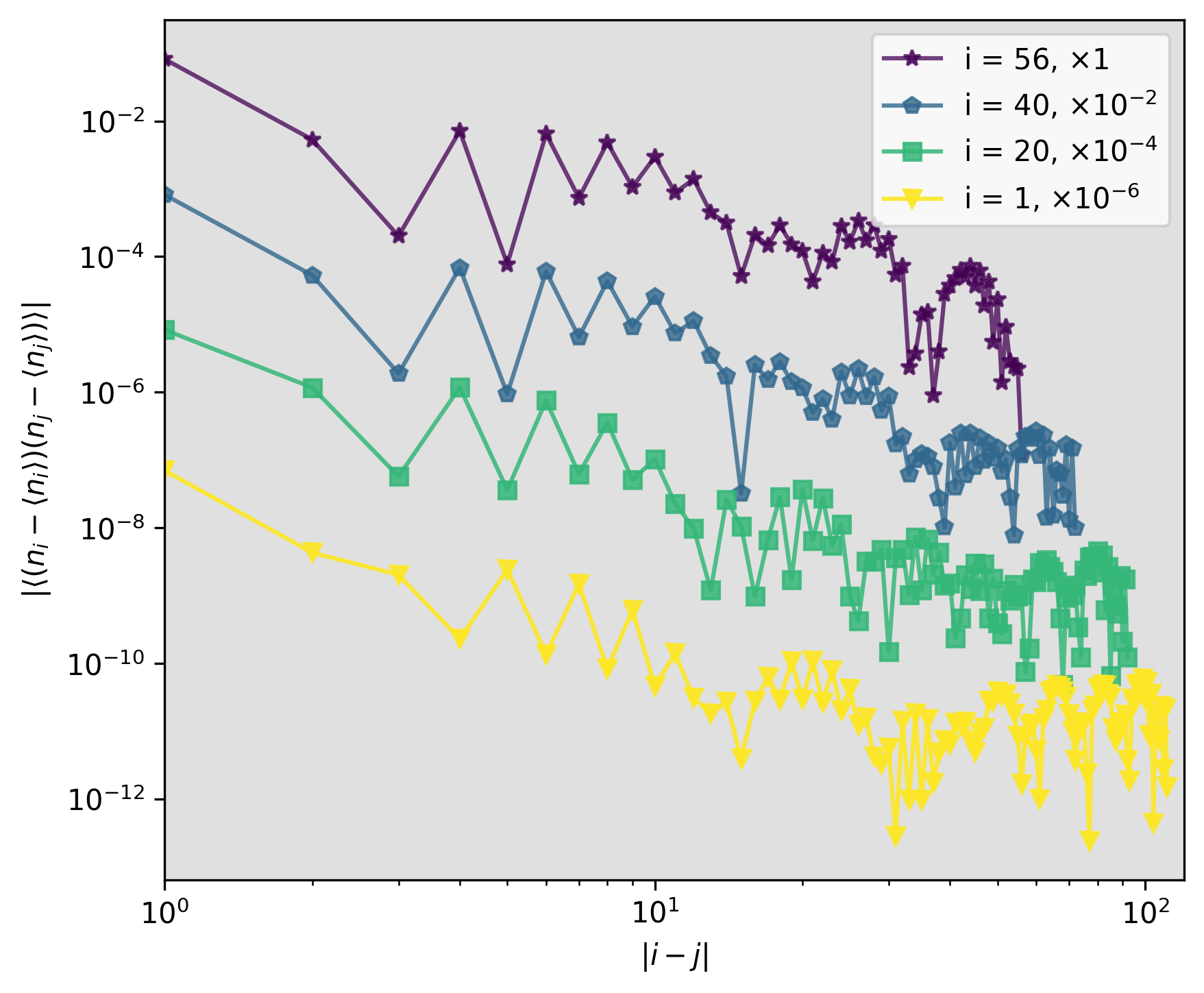}
\caption{Charge density correlation functions $|\braket{(n_i-\braket{n_i})(n_j-\braket{n_j})}|$ with fixed $i$ ($i=1,20,40,56$) for (top) $S=1/2$ and (bottom) $S=3/2$. For clarity, the curves are scaled and shifted.}
\label{fig:nn_finite}
\end{figure}

The difference in the charge densities for $S=1/2, 3/2$ suggests that the density correlation functions may also show distinct behaviors depending on $S$. Figure~\ref{fig:nn_finite} shows $|\braket{(n_i-\braket{n_i})(n_j-\braket{n_j})}|$ for several fixed $i$ ($i=1, 20, 40, 56$) as a function of the other site $j$. For $S=1/2$, the density correlation function simply decays in a power-law fashion for any fixed $i$. For $S=3/2$, however, it shows an upturn behavior at large distance $j\simeq L$ when the site $i$ is close to the other boundary ($i=1, 20$).
Although this may be related to the enhanced $Q,Q''$-oscillations in $\braket{n_j}$ for $S=3/2$, the precise connection is not clear.

\begin{figure}[htb]
\centering
    \includegraphics[width=0.8\columnwidth]{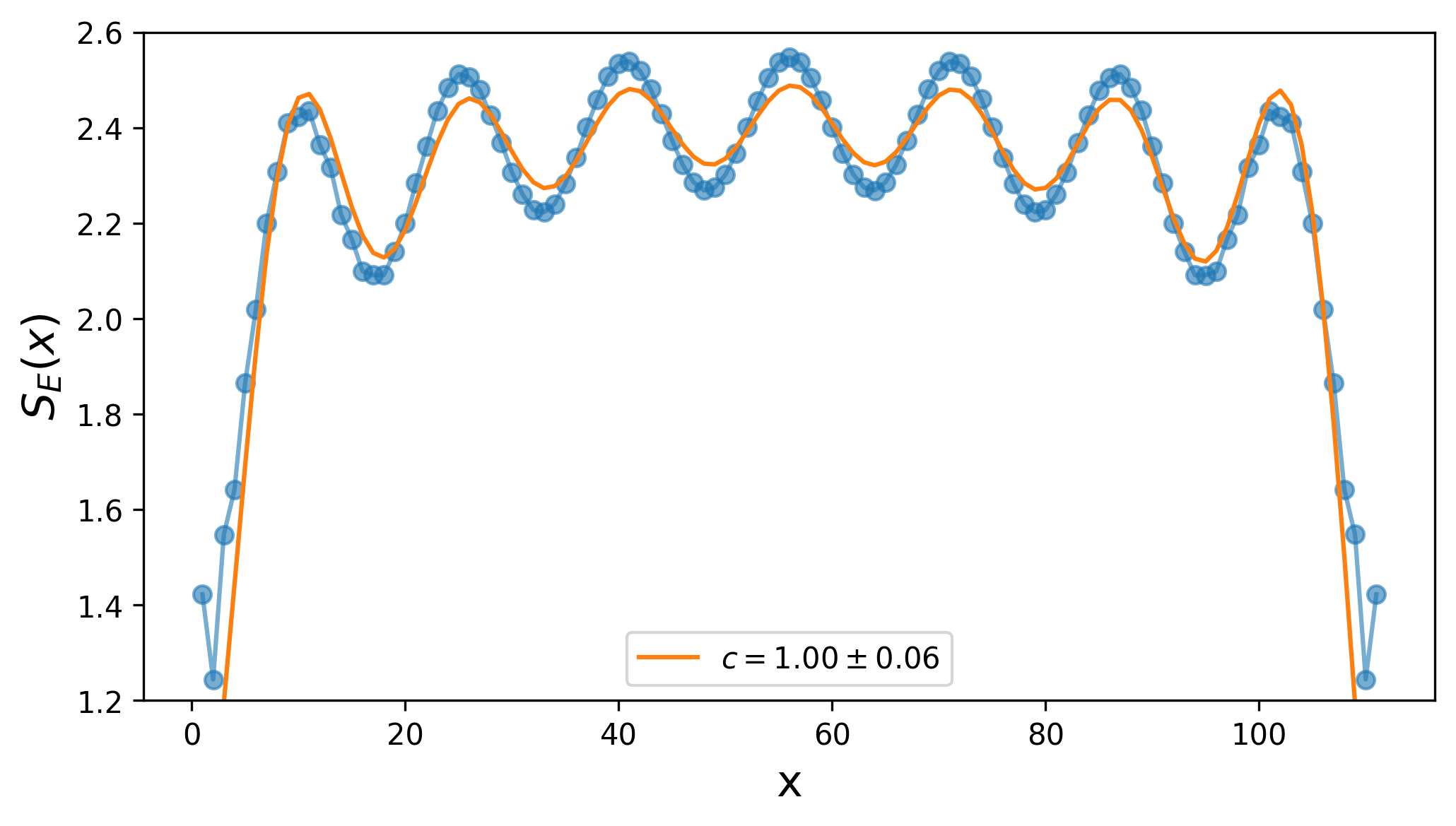}
    \includegraphics[width=0.8\columnwidth]{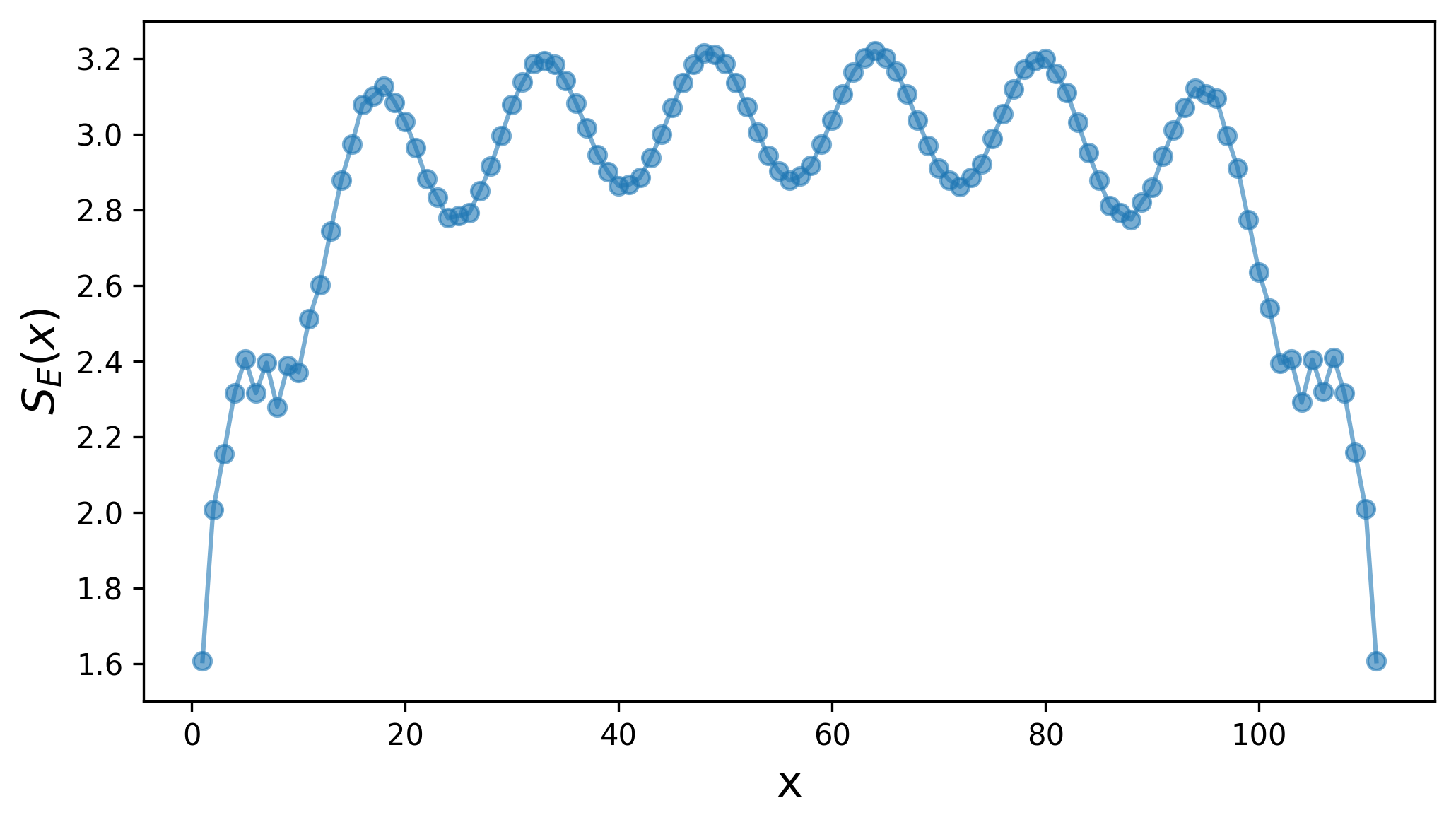}
\caption{Entanglement entropy $S_E$ for (top) $S=1/2$ and (bottom) $S=3/2$. The orange curve for $S=1/2$ shows a fit to the calculation result in the region $2\leq x\leq L-3$.}
\label{fig:EE_finite}
\end{figure}

The anomalous boundary behaviors also affect the entanglement entropy $S_E$ as shown in Fig.~\ref{fig:EE_finite}. For a subsystem with the length $x\gg 1$, the entanglement entropy is described as
\begin{align}
S_E(x)=\frac{c}{6}\log\left(\frac{L}{\pi}\sin\left(\frac{\pi x}{L}\right)\right)+S_0+S_{\rm osc}(x),
\label{eq:EE_finite}
\end{align}
where $S_0$ is a constant and $S_{\rm osc}$ is a boundary contribution~\cite{CalabreseCardy2004,Affleck2006}. The boundary term is written as $S_{\rm osc}(x)=d\sin(qx)/[(L/\pi)\sin(\pi x/L)]$ with the fitting parameters $d,q$. For $S=1/2$, the overall behavior of the numerical result is well described by the formula~\eqref{eq:EE_finite}, which leads to the central charge $c\simeq 1.00$ and the oscillation parameter $q\simeq\pi/8=-2k_F^*$ (mod $2\pi$). The obtained $c$ is consistent with the one evaluated by the iDMRG calculation in the previous section, supporting the validity of our analyses. On the other hand, for $S=3/2$, we cannot obtain a reasonable fit by using the formula ~\eqref{eq:EE_finite} mainly due to a $\pi$-phase shift in the bulk and strong boundary effects. Indeed, the oscillation period of $S_E$ in the bulk region is $\lambda=2\pi/q\simeq 16$, but there is a $\pi$-phase shift compared to that for $S=1/2$ and the oscillation structure is strongly modified around the boundary.

Although the boundary effects discussed above make it difficult to identify the intrinsic bulk hierarchy of correlations from finite open chains alone, they may themselves contain useful information about the nontrivial nature of the bulk state. In the present system, the bulk already exhibits an unusual combination of dominant PDW correlations, reconstructed single-particle structure, and nontrivial charge modes. From this viewpoint, the strong boundary-induced density modulations and the enhancement/suppression of pairing correlations near the edges may not be merely accidental finite-size effects, but could be related to the underlying correlated bulk structure. Clarifying the origin of these boundary effects, and more generally elucidating the relation between bulk reconstruction and boundary response in the Kondo-Heisenberg chain, will therefore be an interesting subject for future study.

\section{summary and discussion}
\label{sec:summary}

In summary, we have shown that the spin-gapped superconducting phase of the one-dimensional Kondo-Heisenberg model can be understood as an interior-gap PDW state generated by strong correlations. By means of iDMRG directly in the thermodynamic limit, we found that the bond-singlet-pairing PDW correlation is the dominant bulk quasi-long-range order for both $S=1/2$ and $S=3/2$. At the same time, the single-particle momentum distribution function $n(k)$ exhibits a reconstructed profile characteristic of interior-gap physics, with two small-Fermi-surface structures at $k_{F1}$ and $k_{F2}=Q-k_F$. In particular, the weak hump-like feature for $S=1/2$ evolves into a clear dip for $S=3/2$, strongly supporting the interior-gap interpretation.

Our results also clarify that this reconstructed structure is qualitatively different from the conventional interior-gap scenario based on mismatched preexisting Fermi surfaces. In the present system, the bare conduction electrons have only a single Fermi surface, and the additional singularity at $k_{F2}$ emerges dynamically through the Kondo coupling together with the dominant PDW correlation. Furthermore, for $S=1/2$, the coexistence of two Fermi-wavevector singularities with central charge $c=1$ indicates that these two fermionic excitations are not independent low-energy modes. A useful phenomenological picture is that the PDW order, the emergent $k_{F2}$ singularity, and the charge-neutral mode at $Q=\pi$ are intertwined manifestations of the same correlated phase, rather than separate ingredients connected by a simple cause-and-effect relation. From a broader perspective, the present results also connect the PDW state in the Kondo-Heisenberg chain to the pocket-like quasiparticle structures discussed previously in the PDW literature within the mean-field approximations, while showing that such physics can arise here in a strongly correlated one-dimensional setting.

Finally, finite DMRG calculations on open chains demonstrate that boundary effects can substantially modify real-space correlations and obscure the true bulk hierarchy of competing orders. The combination of iDMRG and finite DMRG is therefore essential for distinguishing intrinsic bulk properties from boundary-induced structures in the Kondo-Heisenberg chain. Our results provide a unified bulk picture of PDW order, interaction-generated interior-gap reconstruction, and boundary sensitivity in this strongly correlated one-dimensional system.

An important implication of the present results is that the commonly adopted weak-coupling intuition based on a small Fermi surface is already nontrivial in the regime $J_K \sim J_H$. In earlier discussions of the Kondo-Heisenberg chain, it is natural to start from the original conduction-electron Fermi surface and to regard the PDW state as being built on top of a small-Fermi-surface structure, especially when $J_K$ is weak. Our results show, however, that once the Kondo coupling becomes comparable to the Heisenberg exchange, the low-energy structure is no longer captured solely by that naive starting point. Even though the original $k_F$ singularity remains visible, an additional fermionic singularity emerges at $k_{F2}=Q-k_F$, together with the dominant PDW correlation and the neutral mode at $Q=\pi$. From this viewpoint, the present numerical results do not simply replace the small-Fermi-surface picture by a conventional large-Fermi-surface one, but instead point to a more nontrivial Kondo-driven reconstruction in which the weak-coupling small-Fermi-surface based description ceases to be straightforward already at intermediate coupling.

\begin{acknowledgments}
The DMRG calculations were performed with the use of TeNPy Library \cite{TeNPy2018}.  We are grateful to H. Ueda for the constructive comments on DMRG calculations. Numerical calculations have been done using the facilities of the Supercomputer Center, the Institute for Solid State Physics, the University of Tokyo. This study is supported by JSPS KAKENHI Grants No. 22K03513 and No. 26K06946.

\end{acknowledgments}

\bibliography{ref}


\end{document}